\documentclass[12pt]{iopart}
\usepackage{ifvtex}
\usepackage{ifpdf}
\usepackage[dvips]{graphicx}
\usepackage{graphics}
\usepackage{float}
\usepackage{etex}
\usepackage{iopams}
\usepackage{xcolor}
\usepackage{cite}
\usepackage[utf8]{inputenc}
\usepackage[english]{babel}
\usepackage{amsthm}
\usepackage{amssymb}
\usepackage{setstack}
\usepackage{dsfont}
\usepackage{mathrsfs}
\usepackage{eucal}
\usepackage[normalem]{ulem}
\newcommand{\be}{\begin{equation}}
\newcommand{\ee}{\end{equation}}
\newcommand{\brr}{\begin{eqnarray}}
\newcommand{\err}{\end{eqnarray}}
\newcommand{\nn}{\nonumber}
\newcommand{\bd}{\begin{displaymath}}
\newcommand{\ed}{\end{displaymath}}

\newcommand{\bib}{\bibitem}
\newcommand{\col}{\mathrel{\mathop:}=}
\theoremstyle{definition}
\newtheorem{definition}{Definition}
\theoremstyle{remark}
\newtheorem*{remark}{Remark}
\def\alf{\alpha}
\def\bet{\beta}
\def\gam{\gamma}

\def\lam{\lambda}
\def\om{\omega}
\def\eps{\epsilon}
\def\veps{\varepsilon}
\def\rpar{\right)}
\def\lpar{\left(}
\def\rbk{\right]}
\def\lbk{\left[}
\def\rbr{\right\}}
\def\lbr{\left\{}
\def\lb{\label}
\def\im{{\rm i}}

\def\tr{\mbox{${\rm Tr}$}}
\def\ro{\mbox{$\hat{\rho}$}}

\def\rg{\rangle}
\def\lg{\langle}
\def\half{\frac{1}{2}}
\def\lb{\label}
\def\ftn{\footnote}
\begin{document}
\title[On the discrete Wigner function for $\mathrm{SU(N)}$]{On the discrete Wigner function for $\mathbf{SU(N)}$}

\author{Marcelo A. Marchiolli$^{1}$\footnote{Author to whom any correspondence should be addressed.} and 
Di\'{o}genes Galetti$^{2}$}

\address{$^{1}$ Avenida General Os\'{o}rio 414, centro, 14.870-100 Jaboticabal, SP, Brazil}
\address{$^{2}$ Instituto de F\'{\i}sica Te\'{o}rica, Universidade Estadual Paulista, Rua Dr. Bento Teobaldo Ferraz 271, Bloco II,
Barra Funda, 01140-070 S\~{a}o Paulo, SP, Brazil}
 
\ead{\textcolor{blue}{marcelo$\_$march@bol.com.br} and \textcolor{blue}{diogaletti@hotmail.com}}

\begin{abstract}
We present a self-consistent theoretical framework for finite-dimensional discrete phase spaces that leads us to establish a
well-grounded mapping scheme between Schwinger unitary operators and generators of the special unitary group $\mathrm{SU(N)}$.
This general mathematical construction provides a sound pathway to the formulation of a genuinely discrete Wigner function for
arbitrary quantum systems described by finite-dimensional state vector spaces. To illustrate our results, we obtain a general
discrete Wigner function for the group $\mathrm{SU(3)}$ and apply this to the study of a particular three-level system. Moreover,
we also discuss possible extensions to the discrete Husimi and Glauber-Sudarshan functions, as well as future investigations on 
multipartite quantum states. 
\end{abstract}


\vspace{2pc}
Journal-ref: \jpa \textbf{52} (2019) 405305

DOI link: \texttt{https://doi.org/10.1088/1751-8121/ab3bab}
\maketitle
\section{Introduction}

Recent advances in theoretical and experimental investigations consolidate the Wigner function as a fundamental mathematical tool
of wide-ranging use in several branches of physics, chemistry, and engineering \cite{WF2018}. Originally formulated in 1932 by
Eugene Paul Wigner \cite{Wigner1932}, this function represents a historical landmark in the continuous phase-space representations
of quantum mechanics \cite{ZFC2005,CFZ2014}. Despite this enormous success, it is important to emphasize that its counterpart in
the finite-dimensional discrete phase-space representations has been the subject of sound theoretical approaches dating back a few
decades \cite{Fer2011}. In this connection, Tilma, Everitt and coworkers \cite{TESMN} recently proposed a general approach for
constructing Wigner functions associated with continuous representations (Euler angles) that allows us to describe arbitrary spin 
systems. In particular, this kind of construction makes explicit use of symmetries related to the special unitary group 
$\mathrm{SU(N)}$, although it produces Wigner functions that are difficult to visualize for multipartite systems. This apparent 
disadvantage can be circumvented by means of a theoretical formalism that encompasses the generators of $\mathrm{SU(N)}$ and their 
representatives in the finite-dimensional discrete phase spaces, producing, as a by-product, discrete Wigner functions that 
describe the multipartite quantum states characterized by finite state vector spaces.

Nowadays, it is noticeable the efforts developed by some authors in the past to establish a quantum-algebraic framework for
finite-dimensional discrete phase spaces that allows, within other things, to properly describe the quasiprobability distribution
functions in complete analogy with their continuous counterparts, as well as in different scenarios of admissible applications in
quantum mechanics \cite{Woo1987,GP1988,Woo1989,AG1990,GP1992,KP1994,GM1996,Vourdas1996,RG2000,RG2002,Vourdas2003,GHW2004,RMG2005,
Galvao,MRG2005,KM2005,KMR2006,MSG2009,KMS2009,CGV2011,CD2012,MR2012,Van2013,MGD2013,SR2016,SR2017,MK2017,KRG2017,SR2018,KL2018,
WHHH2019}. The mail goal of this paper is to fulfil the aforementioned gap through a self-consistent theoretical framework for
finite-dimensional discrete phase spaces that permits to implement a well-succeeded mapping scheme between generators of the
special unitary group and Schwinger unitary operators. This one-to-one correspondence represents a real twofold-gain for our 
purposes, since it leads us to establish the discrete representatives of these generators in such phase spaces (by means of the 
$\mathit{mod(N)}$-invariant unitary operator basis \cite{GP1992}) and to obtain a general expression for the discrete Wigner
function -- here related to $\mathrm{SU(N)}$ -- which describes arbitrary finite quantum systems. As a consequence, the first
standard examples that emerge from this general formalism take into account the groups $\mathrm{SU(2)}$ and $\mathrm{SU(3)}$, with
special emphasis on the respective discrete Wigner functions. The study of a particular three-level system serves, in this case,
as a toy model employed to illustrate the discrete Wigner function for $\mathrm{SU(3)}$ and its possible implications in physics.
To conclude, we also discuss some relevant points associated with possible extensions to the discrete Husimi and Glauber-Sudarshan
quasiprobability distribution functions, arbitrary spin systems, and multipartite quantum states.

This paper is structured as follows. In Section \ref{s2}, we fix a preliminary mathematical background on finite-dimensional 
discrete phase spaces that leads us to establish a first definition of a discrete Wigner function via $\mathit{mod(N)}$-invariant
unitary operator basis, and also general expressions for mean values. In Section \ref{s3}, we briefly review the basic aspects
of the group $\mathrm{SU(N)}$ with focus on the complete orthonormal operator basis formed by its $N^{2}-1$ generators. Next, we
show the formal connection with the finite-dimensional discrete phase spaces through a well-grounded mapping scheme that allows 
to construct straight relations between generators of the special unitary group and Schwinger unitary operators, as well as to 
determine the discrete representatives of these generators in such phase spaces. This section culminates by presenting an elegant
mathematical expression for the discrete Wigner function -- now associated with the group $\mathrm{SU(N)}$ -- which allows us to
investigate arbitrary quantum systems described by finite-dimensional state vector spaces. Section \ref{s4} is dedicated to
illustrate our results through the group $\mathrm{SU(3)}$, where the corresponding discrete Wigner function is used to investigate
a particular three-level system. To conclude, Section \ref{s5} contains an interesting discussion on possible extensions to the 
discrete Husimi and Glauber-Sudarshan functions, as well as future applications on multipartite quantum states. Two mathematical
appendices concerned with important topics and calculational details of certain expressions used in the previous sections were
added: \ref{apa} shows, in particular, how the parity operator is connected with the $\mathit{mod(N)}$-invariant operator basis 
in the continuum limit; while \ref{apb} exhibits the relations between Gell-Mann and Schwinger unitary operators, and their
corresponding mapped expressions in the finite-dimensional discrete phase spaces.

\section{Finite-dimensional discrete phase spaces}
\lb{s2}

Let us begin with establishing the mathematical prerequisites related to the Schwinger unitary operators and its corresponding
symmetrized basis \cite{GP1988}, whose operational aspects allow to build an effective self-consistent theoretical framework for
the $\mathit{mod}(N)$-invariant unitary operator basis \cite{GP1992}, and subsequently, for the discrete Wigner function
\cite{GM1996}. The sentence ``\textit{finite-dimensional discrete phase space}" means, henceforth, a finite mesh with $N^{2}$
points labelled by discrete variables.

\begin{definition}[Schwinger]
Let $\hat{U}$ and $\hat{V}$ describe a pair of unitary operators defined in a $N$-dimensional state vector space, as well as 
$\{ | u_{\alf} \rg, | v_{\bet} \rg \}$ denote their respective orthonormal eigenvectors related by the symmetrical finite Fourier
kernel (and/or inner product) $\lg u_{\alf} | v_{\bet} \rg = \frac{1}{\sqrt{N}} \om^{\alf \bet}$ with $\om \col \exp \lpar
\frac{2 \pi \im}{N} \rpar$. The general properties \cite{Schwin-book}\ftn{Further results and properties associated with $\hat{U}$
and $\hat{V}$ can be found in Ref. \cite{GM1996}.}
\brr
& & \hat{U}^{\eta} | u_{\alf} \rg = \om^{\alf \eta} | u_{\alf} \rg , \;\; \hat{V}^{\xi} | v_{\bet} \rg = \om^{\bet \xi} 
| v_{\bet} \rg , \nn \\
& & \hat{U}^{\eta} | v_{\bet} \rg = | v_{\bet + \eta} \rg , \;\; \hat{V}^{\xi} | u_{\alf} \rg = | u_{\alf - \xi} \rg , \nn \\
& & \hat{U}^{N} = \hat{\mathds{I}} , \;\; \hat{V}^{N} = \hat{\mathds{I}} , \;\; \hat{V}^{\xi} \hat{U}^{\eta} = \om^{\eta \xi}
\hat{U}^{\eta} \hat{V}^{\xi} , \nn 
\err
where $\hat{\mathds{I}}$ corresponds to the identity operator, consist of fundamental basic mathematical rules that characterize
the aforementioned unitary operators. Besides, the discrete labels $\{ \alf,\bet,\eta,\xi \}$ obey the arithmetic modulo $N$.
\end{definition}

\begin{remark}
The commutation relation $[ \hat{U},\hat{V} ] = (1 - \om) \hat{U} \hat{V}$ for Schwinger unitary operators and its complementary
property $\hat{V} \hat{U} = \om \hat{U} \hat{V}$, lead us to establish some related additional results as follow:
\brr
& & \bigl[ \hat{U}, \bigl[ \hat{U}, \bigl[ \hat{U}, \ldots [ \hat{U},\hat{V} ] \ldots \bigr] \bigr] \bigr] = ( 1 - \om )^{p}
\hat{U}^{p} \hat{V} , \nn \\
& & \bigl[ \hat{V}, \bigl[ \hat{V}, \bigl[ \hat{V}, \ldots [ \hat{V},\hat{U} ] \ldots \bigr] \bigr] \bigr] = 
( 1 - \om^{\ast} )^{p} \hat{V}^{p} \hat{U} , \nn \\
& & \bigl( \hat{U} \hat{V} \bigr)^{q} = \om^{\half q(q-1)} \hat{U}^{q} \hat{V}^{q} , \nn
\err
with $p,q \in \mathbb{N}^{\ast}$. On the other hand, such results permit to infer that exists an uncertainty relation underlying
the operators $\hat{U}$ and $\hat{V}$. In the recent past \cite{MM2013}, it was introduced a quantum-algebraic framework embracing
a new uncertainty principle for these unitary operators that generalizes and strengthens the Massar-Spindel inequality 
\cite{MS2008} -- this inequality determines, in particular, a new set of restrictions upon the selective process of signals and
wavelet bases. Within this scope, it is worth mentioning that Bagchi and Pati \cite{BP2016} also derived new uncertainty relations
for arbitrary unitary operators acting on finite-dimensional state vector spaces, whose respective tighter bounds were obtained
for different situations. Recently, it was shown that minimum-uncertainty states saturate the Massar-Spindel inequality 
\cite{HHH2019}.
\end{remark}

\begin{definition}[Schwinger] Let us introduce the set of $N^{2}$ operators
\be
\lb{eq1s2}
\hat{S}_{\mathrm{S}}(\eta,\xi) = \frac{1}{\sqrt{N}} \om^{\half \eta \xi} \hat{U}^{\eta} \hat{V}^{\xi} \qquad
(\eta,\xi = 0,\ldots,N-1) ,
\ee
which represents a symmetrized version of the unitary operator basis $\hat{S}(\eta,\xi) = \frac{1}{\sqrt{N}} \hat{U}^{\eta}
\hat{V}^{\xi}$ originally proposed by Schwinger \cite{Schwin-book}. The labels $\eta$ and $\xi$ are associated with the discrete 
dual variables of an $N^{2}$-dimensional phase space; besides, it can be also verified by direct inspection that 
$\hat{S}_{\mathrm{S}}(\eta,\xi)$ is invariant under the changes $\hat{U} \rightarrow \hat{V}$ and $\hat{V} \rightarrow 
\hat{U}^{-1}$, followed by $\eta \rightarrow \xi$ and $\xi \rightarrow - \eta$ (this result depicts the pre-symplectic character 
of the symmetrized unitary operator basis \cite{GP1988,AG1990}). Additionally, note that both the inverse element
\bd 
\hat{S}_{\mathrm{S}}^{-1}(\eta,\xi) = \hat{S}_{\mathrm{S}}^{\dagger}(\eta,\xi) = \hat{S}_{\mathrm{S}}(-\eta,-\xi)
\ed
and the similarity transformation
\be
\lb{ext-1}
\lbk \sqrt{N} \hat{S}_{\mathrm{S}}(\alf,-\bet) \rbk \hat{S}_{\mathrm{S}}(\eta,\xi) \lbk \sqrt{N} \hat{S}_{\mathrm{S}}^{\dagger}
(\alf,-\bet) \rbk = \om^{-( \bet \eta + \alf \xi )} \hat{S}_{\mathrm{S}}(\eta,\xi)
\ee
represent good examples of relevant mathematical properties associated with $\hat{S}_{\mathrm{S}}(\eta,\xi)$ -- see Ref. 
\cite{GM1996} for complementary results. Summarising, $\{ \hat{S}_{\mathrm{S}}(\eta,\xi) \}_{\eta,\xi=0,\ldots,N-1}$ constitutes a
complete orthonormal operator basis that allows us to determine all possible dynamical quantities belonging to the physical system
under investigation; consequently, any linear operator $\hat{O}$ can be decomposed in this basis as\ftn{It is worth noting that
$\Tr \bigl[ \hat{S}_{\mathrm{S}}^{\dagger}(\eta^{\prime},\xi^{\prime}) \hat{S}_{\mathrm{S}}(\eta,\xi) \bigr] = 
\delta_{\eta,\eta^{\prime}}^{[N]} \delta_{\xi,\xi^{\prime}}^{[N]}$ establishes a fundamental property since it ensures that such a
decomposition is unique. The superscript $[N]$ on the Kronecker deltas denotes that this function is different from zero when its
labels are $\mathit{mod}(N)$-congruent.}
\be
\lb{eq2s2}
\hat{O} = \sum_{\eta,\xi=0}^{N-1} \mathcal{O}(\eta,\xi) \hat{S}_{\mathrm{S}}(\eta,\xi) ,
\ee
where the coefficients $\mathcal{O}(\eta,\xi)$ are given by $\Tr \bigl[ \hat{S}_{\mathrm{S}}^{\dagger}(\eta,\xi) \hat{O} \bigr]$.
When $\hat{O} \equiv \hat{\rho}$, these coefficients describe the discrete Wigner characteristic function $\chi_{\mathtt{W}}
(\eta,\xi) \col \Tr \bigl[ \hat{S}_{\mathrm{S}}^{\dagger}(\eta,\xi) \hat{\rho} \bigr]$.
\end{definition}

\begin{definition}[Galetti--de Toledo Piza] The $\mathit{mod}(N)$-invariant unitary operator basis was originally defined as the
discrete Fourier transform of the symmetrized basis \cite{GP1992}
\be
\lb{eq3s2}
\hat{G}(\mu,\nu) = \frac{1}{\sqrt{N}} \sum_{\eta,\xi=0}^{N-1} \om^{-(\mu \eta + \nu \xi)} \om^{\half N \Phi(\eta,\xi;N)}
\hat{S}_{\mathrm{S}}(\eta,\xi) ,
\ee
being the additional phase $\Phi(\eta,\xi;N) \col N \mathit{I}_{\eta}^{N} \mathit{I}_{\xi}^{N} - \eta \mathit{I}_{\xi}^{N} - \xi
\mathit{I}_{\eta}^{N}$ responsible for the $\mathit{mod}(N)$-invariance of this new operator basis \eref{eq3s2}, where 
$\mathit{I}_{\varepsilon}^{N} = \lbk \frac{\varepsilon}{N} \rbk$ corresponds to the integer part of $\varepsilon$ with respect to 
$N$. Such an additional phase is irrelevant in describing single operators, however, it may be meaningful for products of 
operators and/or composition laws \cite{GM1996}.\ftn{In fact, the term $\om^{\half N \Phi(\eta,\xi;N)}$ can be suppressed when one
deals with discrete labels obeying the arithmetic modulo $N$.} By analogy with \Eref{eq2s2}, the decomposition of any linear 
operator
\be
\lb{eq4s2}
\hat{O} = \frac{1}{N} \sum_{\mu,\nu=0}^{N-1} \mathrm{O}(\mu,\nu) \hat{G}(\mu,\nu)
\ee
can also be verified in such a case, with $\mathrm{O}(\mu,\nu) = \Tr \bigl[ \hat{G}^{\dagger}(\mu,\nu) \hat{O} \bigr]$ exhibiting
a one-to-one correspondence between operators and functions belonging to an $N^{2}$-dimensional phase space characterized by the
discrete labels $\mu$ and $\nu$. In particular, these coefficients allow us to define, for $\hat{O} \equiv \hat{\rho}$, the
discrete Wigner function $\mathit{W}(\mu,\nu) \col \Tr \bigl[ \hat{G}^{\dagger}(\mu,\nu) \hat{\rho} \bigr]$.
\end{definition}

\begin{remark}
There is an immediate link between both the coefficients $\mathrm{O}(\mu,\nu)$ and $\mathcal{O}(\eta,\xi)$, given by the
discrete Fourier transform
\be
\lb{eq5s2}
\mathrm{O}(\mu,\nu) = \frac{1}{\sqrt{N}} \sum_{\eta,\xi=0}^{N-1} \om^{\mu \eta + \nu \xi} \om^{- \half N \Phi(\eta,\xi;N)}
\mathcal{O}(\eta,\xi) .
\ee
\Eref{eq5s2} exhibits essentially the mathematical role of the discrete Fourier transform in connecting the $N^{2}$-dimensional
discrete phase spaces described by the pair $(\mu,\nu)$ and its respective dual pair $(\eta,\xi)$; then, it becomes quite easy
to show the connection between discrete Wigner function and its respective characteristic function, that is
\be
\lb{eq6s2}
\mathit{W}(\mu,\nu) = \frac{1}{\sqrt{N}} \sum_{\eta,\xi=0}^{N-1} \om^{\mu \eta + \nu \xi} \om^{- \half N \Phi(\eta,\xi;N)}
\chi_{\mathtt{W}}(\eta,\xi) ,
\ee
in complete analogy to the continuous case \cite{Orszag-book}. Summarizing, the functions $\mathrm{O}(\mu,\nu)$ and
$\mathit{W}(\mu,\nu)$ correspond, in this context, to a well-established one-to-one mapping between operators and functions
embedded in a finite phase space characterized by the discrete variables $\mu$ and $\nu$.\ftn{In particular, when one deals with
discrete momentum and coordinate-like variables, these variables assume integer values in the symmetric interval $[-\ell,\ell]$
for $\ell = \frac{N-1}{2}$ fixed and a given odd $N$ -- see, for instance, Ref. \cite{GM1996} for description of discrete coherent
states.} Note that mappings of Hermitian operators in the $N^{2}$-dimensional discrete phase space lead us to obtain real
functions.
\end{remark}

Next, we present an important property related to the trace of the product of two bounded operators $\hat{A}$ and $\hat{B}$,
\be
\lb{eq7s2}
\Tr [ \hat{A} \hat{B} ] = \frac{1}{N} \sum_{\mu,\nu=0}^{N-1} \mathrm{A}(\mu,\nu) \mathrm{B}(\mu,\nu) = \sum_{\eta,\xi=0}^{N-1}
\mathcal{A}(\eta,\xi) \mathcal{B}(-\eta,-\xi) ,
\ee
which corresponds to the overlap of their respective mappings in the $G$ or $S_{\mathrm{S}}$ bases; in addition, if one considers
the density operators $\hat{\rho}_{1}$ and $\hat{\rho}_{2}$, it turns immediate to establish that
\be
\lb{eq8s2}
\fl \qquad \Tr [ \hat{\rho}_{1} \hat{\rho}_{2} ] = \frac{1}{N} \sum_{\mu,\nu=0}^{N-1} \mathit{W}_{1}(\mu,\nu) 
\mathit{W}_{2}(\mu,\nu) = \sum_{\eta,\xi=0}^{N-1} \chi_{\mathtt{W},1}(\eta,\xi) \chi_{\mathtt{W},2}(-\eta,-\xi) .
\ee
Similarly, an expression for the mean value $\lg \hat{O} \rg \equiv \Tr [ \hat{O} \hat{\rho} ]$ can also be obtained from this
property,
\be
\lb{eq9s2}
\lg \hat{O} \rg = \frac{1}{N} \sum_{\mu,\nu=0}^{N-1} \mathrm{O}(\mu,\nu) \mathit{W}(\mu,\nu) = \sum_{\eta,\xi=0}^{N-1}
\mathcal{O}(\eta,\xi) \chi_{\mathtt{W}}(-\eta,-\xi) .
\ee
So, all the relevant quantities needed to describe the kinematical/dynamical content of physical systems with finite
state spaces have, in this description of quantum mechanics, their representatives in the finite-dimensional discrete phase space.
A few years ago, by means of the extended Cahill-Glauber formalism for finite-dimensional spaces \cite{RMG2005,MRG2005}, the
aforementioned results were generalized in order to include the discrete Husimi and Glauber-Sudarshan functions. 

Now, let us say a few words about $\hat{G}(\mu,\nu)$ and $\hat{S}_{\mathrm{S}}(\eta,\xi)$, since both the discrete bases are
connected through the \Eref{eq3s2}. There are other proposals of discrete bases for finite-dimensional phase spaces in literature,
with convenient inherent mathematical properties, which can also be applied in analogous quantum systems \cite{Fer2011}. In 
particular, Klimov and co-workers \cite{KM2005,KMS2009} proposed equivalent mathematical expressions of discrete bases for 
finite state spaces, where they basically showed that discrete Wigner functions depend on the specific phase choice for such
bases. So, in Appendix A we present certain interesting aspects of those aforementioned discrete bases and their consequences on 
the different definitions of discrete Wigner functions. 

\section{Mappings in $\mathbf{SU(N)}$}
\lb{s3}

After establishing a self-consistent theoretical framework for finite-dimensional discrete phase spaces, let us now implement, at
this moment, an important set of mathematical and physical results involving the elements of the special unitary group
$\mathrm{SU(N)}$ and the Schwinger unitary operators. In particular, we will show how the generators of the Lie algebra $su(N)$
can be mapped upon $N^{2}$-dimensional discrete phase-space representatives through the use of $\mathit{mod}(N)$-invariant unitary 
operator basis; consequently, as a relevant by-product of this biunivocal mapping, a discrete Wigner function for $N$-level 
systems associated with the Hilbert space $\mathcal{H}_{N}$ can be properly obtained.

\begin{definition}
Let $\hat{g}_{i}$ with $i=1,\ldots,N^{2}-1$ denote the generators of the Lie algebra $su(N)$ characterized by $N \times N$
skew-Hermitian matrices obeying the relations $\hat{g}_{i}^{\dagger} = \hat{g}_{i}$, $\Tr [ \hat{g}_{i} ] = 0$, and 
$\Tr [ \hat{g}_{i} \hat{g}_{j} ] = 2 \delta_{ij}$. In addition, the structure constants \cite{CS1978}
\bd
\mathscr{F}_{ijk} = - \frac{\im}{4} \Tr \bigl[ [ \hat{g}_{i},\hat{g}_{j} ] \hat{g}_{k} \bigr] \qquad \mbox{(antisymmetric tensor)}
\ed
of this algebra are associated with the commutation relation
\bd
[ \hat{g}_{i},\hat{g}_{j} ] = 2 \im \sum_{k=1}^{N^{2}-1} \mathscr{F}_{ijk} \hat{g}_{k} \, ,
\ed
whereas the anti-structure constants 
\bd
\mathscr{D}_{ijk} = \frac{1}{4} \Tr \bigl[ \{ \hat{g}_{i},\hat{g}_{j} \} \hat{g}_{k} \bigr] \qquad \mbox{(symmetric tensor)}
\ed
are related to the anticommutation relation
\bd
\{ \hat{g}_{i},\hat{g}_{j} \} = \frac{4}{N} \delta_{ij} \hat{\mathds{I}}_{N} + 2 \sum_{k=1}^{N^{2}-1} \mathscr{D}_{ijk} 
\hat{g}_{k} \, , 
\ed
where $\hat{\mathds{I}}_{N}$ represents the $N$-dimensional unit matrix. Note that both the constants $\mathscr{F}_{ijk}$ and 
$\mathscr{D}_{ijk}$ can be found in tabulated form in the literature for different values of $N$ \cite{MW-book,AL-book}. For 
the sake of clarity, let us just briefly mention that $su(2)$ has as generating operators the Pauli matrices $\hat{\sigma}_{i}$
for $i=x,y,z$ with $\mathscr{F}_{ijk} = \eps_{ijk}$ (Levi-Civita symbol) and $\mathscr{D}_{ijk} = 0$.
\end{definition}

\begin{remark}
To construct systematically the generators $\{ \hat{g}_{i} \}_{i=1,\ldots,N^{2}-1}$ of the group $\mathrm{SU(N)}$, let us
initially define two Hermitian operators \cite{MW-book}
\bd
\fl \quad \hat{\mathscr{U}}_{\alf,\bet} \col \hat{\mathscr{P}}_{\alf,\bet} + \hat{\mathscr{P}}_{\bet,\alf} \;\; \mbox{and} \;\;
\hat{\mathscr{V}}_{\alf,\bet} \col - \im \bigl( \hat{\mathscr{P}}_{\alf,\bet} - \hat{\mathscr{P}}_{\bet,\alf} \bigr) \qquad
(0 \leq \alf < \bet \leq N-1)
\ed
which are specific combinations of the transition operators $\hat{\mathscr{P}}_{\alf,\bet} = | \alf \rg \lg \bet |$; in both
cases, $| \alf \rg$ and $| \bet \rg$ denote complete orthonormal bases in $\mathcal{H}_{N}$. In addition, let us also introduce 
the Hermitian operator
\bd
\fl \quad \hat{\mathscr{W}}_{\gam} = \sqrt{\frac{2}{(\gam + 1)(\gam + 2)}} \lbk \sum_{\sigma = 0}^{\gam} 
\hat{\mathscr{P}}_{\sigma,\sigma} - (\gam + 1) \hat{\mathscr{P}}_{\gam + 1,\gam + 1} \rbk \qquad (0 \leq \gam \leq N-2) ,
\ed
where $\hat{\mathscr{P}}_{\sigma,\sigma} = | \sigma \rg \lg \sigma |$ corresponds to the projection operators. In this way, the
set formed by the $N^{2}-1$ orthogonal operators 
\bd
\fl \qquad \{ \hat{g} \} = \lbr \hat{\mathscr{U}}_{0,1},\hat{\mathscr{U}}_{0,2},\ldots,\hat{\mathscr{U}}_{1,2},\ldots,
\hat{\mathscr{V}}_{0,1},\hat{\mathscr{V}}_{0,2},\ldots,\hat{\mathscr{V}}_{1,2},\ldots,\hat{\mathscr{W}}_{0},
\hat{\mathscr{W}}_{1},\ldots,\hat{\mathscr{W}}_{N-2} \rbr
\ed
completely characterizes the generators of $\mathrm{SU(N)}$ -- indeed, it corresponds to a complete orthonormal operator 
basis \cite{HE1981}.
\end{remark}

In this general algebraic-theoretical approach, the decomposition of any linear operator $\hat{O}$ is expressed as follows:
\be
\lb{eq10s3}
\hat{O} = \frac{1}{N} \Tr [ \hat{O} ] \hat{\mathds{I}}_{N} + \half \sum_{i=1}^{N^{2}-1} \mathcal{O}_{i} \hat{g}_{i} \, ,
\ee
where the coefficients $\mathcal{O}_{i}$ are given by $\Tr [ \hat{g}_{i} \hat{O} ]$. So, for $N$-level quantum systems related to
the Hilbert space $\mathcal{H}_{N}$, the associated density operator $\hat{\rho}$ can be promptly determined from their $N^{2}-1$
mean values $\lg \hat{g}_{i} \rg = \Tr [ \hat{g}_{i} \hat{\rho} ]$, namely, 
\be
\lb{eq11s3}
\hat{\rho} = \frac{1}{N} \hat{\mathds{I}}_{N} + \half \sum_{i=1}^{N^{2}-1} \lg \hat{g}_{i} \rg \hat{g}_{i} \, . 
\ee
Now, from the experimental point of view, it is sufficient to measure the components of the Bloch vector $\mathbf{g} = \lpar
\lg \hat{g}_{1} \rg, \ldots, \lg \hat{g}_{N^{2}-1} \rg \rpar \in \mathbb{R}^{N^{2}-1}$ to obtain an acceptable description of 
such states \cite{Ki2003,BK2003}. Next, let us combine Equations (\ref{eq10s3}) and (\ref{eq11s3}) in order to achieve a compact
expression for the mean value
\be
\lb{eq12s3}
\lg \hat{O} \rg = \frac{1}{N} \Tr [ \hat{O} ] + \half \sum_{i=1}^{N^{2}-1} \mathcal{O}_{i} \lg \hat{g}_{i} \rg .
\ee
An immediate extension of this result refers to the product of two operators $\hat{A}$ and $\hat{B}$, that is
\be
\lb{eq13s3}
\lg \hat{A} \hat{B} \rg = \lg \hat{A} \rg \lg \hat{B} \rg + \frac{1}{4} \sum_{i,j=1}^{N^{2}-1} \mathcal{A}_{i} \mathcal{B}_{j}
\lpar \lg \hat{g}_{i} \hat{g}_{j} \rg - \lg \hat{g}_{i} \rg \lg \hat{g}_{j} \rg \rpar ,
\ee
being the second term of the right-hand side responsible for correlations associated with generators and quantum states.

\subsection{Connections with finite-dimensional discrete phase spaces}

To begin with, let $\mathcal{H}_{N}$ describe the $N$-dimensional state vector space \cite{Halmos} previously stated in Section 2.
Therefore, it seems quite reasonable to assume that $| \alf \rg \equiv | u_{\alf} \rg$, which implies that both the transition and
projection operators can be also written as specific combinations of the Schwinger unitary operators, this important connection
being the necessary link to characterize the generators of $\mathrm{SU(N)}$ in finite-dimensional discrete phase spaces.\ftn{In 
fact, such a connection is justified through \Eref{eq4s2} for $\hat{O} \equiv \hat{g}_{i}$ and it corresponds to a change of
basis,
\bd
\hat{g}_{i} = \frac{1}{N} \sum_{\mu,\nu = 0}^{N-1} \Tr [ \hat{G}^{\dagger}(\mu,\nu) \hat{g}_{i} ] \hat{G}(\mu,\nu) \qquad
(i=1,\ldots,N^{2}-1) .
\ed
}

Let us start with the mappings of $\hat{\mathscr{P}}_{\alf,\bet}$ in the $N^{2}$-dimensional discrete phase space, namely,
\bd
\fl \qquad \hat{\mathscr{P}}_{\alf,\bet} = | u_{\alf} \rg \lg u_{\bet} | = \frac{1}{N} \sum_{\eta = 0}^{N-1} \om^{- \eta \alf}
\hat{U}^{\eta} \hat{V}^{\bet - \alf} = \frac{1}{\sqrt{N}} \sum_{\eta = 0}^{N-1} \om^{- \half \eta (\alf + \bet)} 
\hat{S}_{\mathrm{S}}(\eta,\bet - \alf) ,
\ed
as well as the projection operators
\bd
\fl \qquad \hat{\mathscr{P}}_{\sigma,\sigma} = | u_{\sigma} \rg \lg u_{\sigma} | = \frac{1}{N} \sum_{\eta = 0}^{N-1} 
\om^{- \eta \sigma} \hat{U}^{\eta} = \frac{1}{\sqrt{N}} \sum_{\eta = 0}^{N-1} \om^{- \eta \sigma} \hat{S}_{\mathrm{S}}(\eta,0) .
\ed
These particular results enable us to obtain the Hermitian operators
\brr
\lb{eq14s3}
& & \hat{\mathscr{U}}_{\alf,\bet} = \frac{1}{N} \sum_{\eta = 0}^{N-1} \hat{U}^{\eta} \lpar \om^{- \eta \alf} 
\hat{V}^{\bet - \alf} + \om^{- \eta \bet} \hat{V}^{N - (\bet - \alf)} \rpar , \\
\lb{eq15s3}
& & \hat{\mathscr{V}}_{\alf,\bet} = - \frac{\im}{N} \sum_{\eta = 0}^{N-1} \hat{U}^{\eta} \lpar \om^{- \eta \alf} 
\hat{V}^{\bet - \alf} - \om^{- \eta \bet} \hat{V}^{N - (\bet - \alf)} \rpar ,
\err
and
\be
\lb{eq16s3}
\fl \qquad \hat{\mathscr{W}}_{\gam} = \sqrt{\frac{2}{(\gam + 1)(\gam + 2)}} \frac{1}{N} \sum_{\eta = 0}^{N-1} \lbk 
\sum_{\sigma = 0}^{\gam} \om^{- \eta \sigma} - (\gam+1) \om^{- \eta (\gam+1)} \rbk \hat{U}^{\eta}
\ee
as functions of the Schwinger unitary operators, obeying the restrictions imposed on the discrete labels $\alf,\bet,\gam$. Since
$\hat{\mathscr{U}}_{\alf,\bet}$, $\hat{\mathscr{V}}_{\alf,\bet}$, and $\hat{\mathscr{W}}_{\gam}$ are responsible for the
generators of the group $\mathrm{SU(N)}$, their respective mappings represent a sound mathematical framework that leads us to an
alternative description of physical systems in finite-dimensional discrete phase spaces.

For example, let us now consider the well-known group $\mathrm{SU(2)}$ and its corresponding generators $\{ \hat{g} \} = \{
\hat{g}_{1},\hat{g}_{2},\hat{g}_{3} \} = \{ \hat{\mathscr{U}}_{0,1}, \hat{\mathscr{V}}_{0,1}, \hat{\mathscr{W}}_{0} \}$ written
in terms of the unitary operators $\hat{U}$ and $\hat{V}$: $\hat{g}_{1} = \hat{V}$, $\hat{g}_{2} = - \im \hat{U} \hat{V}$, and 
$\hat{g}_{3} = \hat{U}$. The matrix representation of these results in the basis $\{ | u_{0} \rg, | u_{1} \rg \}$ reproduces 
exactly the Pauli matrices $\{ \hat{\sigma}_{i} \}_{i = x,y,z}$ in a one-to-one correspondence, namely, $\hat{\sigma}_{i} =
\hat{V} \delta_{ix} - \im \hat{U} \hat{V} \delta_{iy} + \hat{U} \delta_{iz}$; consequently, the $\mathit{mod}(2)$-invariant
unitary operator basis \eref{eq3s2} achieves the simple form
\be
\lb{ext-2}
\hat{G}(\mu,\nu) = \half \lbk \hat{\mathds{I}}_{2} + (-1)^{\nu} \hat{\sigma}_{x} + (-1)^{\mu + \nu + 1} \hat{\sigma}_{y} +
(-1)^{\mu} \hat{\sigma}_{z} \rbk .
\ee
Next, we briefly define the density-matrix vectorial space for $N$-level systems.

\subsection{The density matrix}

Now, let us characterize the density-matrix (or equivalently, density-operator) space
\bd
\fl \qquad \mathscr{L}_{+,1}(\mathcal{H}_{N}) = \lbr \ro \in \mathscr{L}(\mathcal{H}_{N}) \; | \; \tr [ \ro ] = 1 , \; \ro = 
\ro^{\dagger} , \; \rho_{\ell} \geq 0 \; (\ell = 1,\ldots,N) \rbr
\ed
for $N$-level systems associated with the Hilbert space $\mathcal{H}_{N}$, which exhibits three important basic premises for 
$\ro$: (i) $\Tr [ \ro ] = 1$ (the normalisation condition is preserved), (ii) $\ro = \ro^{\dagger}$ (by definition, $\ro$ consists 
of a Hermitian matrix), and (iii) $\rho_{\ell} \in \mathbb{R}_{+}$ (the eigenvalues are positive). The notation 
$\mathscr{L}(\mathcal{H}_{N})$ corresponds to the set of linear operators on $\mathcal{H}_{N}$ \cite{Ki2003}. Note that
$\Tr [ \ro^{2} ] \leq 1$ can be considered as a further property, but is not necessary under certain circumstances, as the
equality in this situation is reached only for pure states. In fact, this specific property permits us to connect both the 
discrete Wigner function \eref{eq6s2} and Bloch vector, since the relation
\be
\lb{ext-3}
\Tr [ \ro^{2} ] = \frac{1}{N} \sum_{\mu,\nu=0}^{N-1} \mathit{W}^{2}(\mu,\nu) = \frac{1}{N} + \half | \mathbf{g} |^{2} \leq 1
\ee
is always verified. With respect to the set of eigenvalues $\{ \rho_{\ell} \}_{\ell=1,\ldots,N}$, they are basically determined
from the polynomial equation
\bd
\fl \qquad P(\rho) \col \det (\ro - \rho \mathds{I}_{N}) = \rho^{N} - S_{1} \rho^{N-1} + S_{2} \rho^{N-2} + \ldots + 
(-1)^{N} S_{N} = 0 ,
\ed
whose coefficients satisfy the recorrence relation \cite{BK2003}
\bd
S_{r} = \frac{1}{r} \sum_{s=1}^{r} (-1)^{s-1} \Tr [ \ro^{s} ] S_{r-s} \quad (r \geq 2)
\ed
such that $S_{0} \col 1$ and $S_{1} = 1$. This mathematical recipe presents some disadvantages associated with the nontrivial
calculations of terms like $\lpar \Tr [ \ro^{m} ] \rpar^{n}$ -- see Refs. \cite{Ki2003,BK2003} for technical details.

\subsection{The discrete Wigner function}

To determine the discrete Wigner function for $N$-level systems associated with $\mathcal{H}_{N}$, only two results are needed:
the first one corresponds to its definition in the $N^{2}$-dimensional discrete phase space, $\mathit{W}(\mu,\nu) \col \Tr \bigl[
\hat{G}^{\dagger}(\mu,\nu) \hat{\rho} \bigr]$, as stated in Section 2; while the second one refers to the expansion \eref{eq11s3}
of the density operator $\hat{\rho}$, which deals with the generators $\{ \hat{g}_{i} \}_{i=1,\ldots,N^{2}-1}$ of the group
$\mathrm{SU(N)}$. This mixing of remarkable results leads to establish a solid theoretical background that allows us, among other
things, to determine a compact expression for the discrete Wigner function, that is,\ftn{The discrete Wigner characteristic
function also exhibits an analogous form,
\bd
\chi_{\mathtt{W}}(\eta,\xi) = \frac{1}{\sqrt{N}} \delta_{\eta,0}^{[N]} \delta_{\xi,0}^{[N]} + \half \sum_{i=1}^{N^{2}-1}
\lg \hat{g}_{i} \rg \lpar \hat{g}_{i} \rpar \! (\eta,\xi) \qquad (0 \leq \eta,\xi \leq N-1) ,
\ed
with $\lpar \hat{g}_{i} \rpar \! (\eta,\xi)$ referring to the mapped expressions of the generators in the symmetrized unitary
operator basis $\hat{S}_{\mathrm{S}}(\eta,\xi)$ -- see \Eref{eq6s2} for connection with discrete Wigner function.} 
\be
\lb{eq17s3}
\fl \qquad \mathit{W}(\mu,\nu) = \frac{1}{N} + \half \sum_{i=1}^{N^{2}-1} \lg \hat{g}_{i} \rg \lpar \hat{g}_{i} \rpar \! 
(\mu,\nu) \qquad (0 \leq \mu,\nu \leq N-1) 
\ee
where $\lg \hat{g}_{i} \rg$ denotes the components of the Bloch vector and $\lpar \hat{g}_{i} \rpar \! (\mu,\nu)$ corresponds to
its respective mapped expressions in the $\mathit{mod}(N)$-invariant operator basis. Therefore, the protagonism from these
generators turns completely evident in \Eref{eq17s3}. For this reason, the general mapped expressions of the Hermitian operators 
$\hat{\mathscr{U}}_{\alf,\bet}$, $\hat{\mathscr{V}}_{\alf,\bet}$, and $\hat{\mathscr{W}}_{\gam}$ assume, in this context, the
following forms for $N > 2$ (see \ref{apb}):
\brr
\lb{eq18s3}
\eqalign{
\fl \qquad \bigl( \hat{\mathscr{U}}_{\alf,\bet} \bigr)(\mu,\nu) &= 2 \delta_{\mu,\veps}^{[N]} \cos \lbk \frac{2 \pi \nu}{N} 
(\bet - \alf) \rbk \qquad \mbox{if} \;\; \alf + \bet = 2 \veps \;\; \mbox{and} \;\; \veps \in \mathbb{N}^{\ast} , \cr
\fl \qquad &= \frac{2}{N} \frac{\sin \lbk \lpar \mu - \frac{\bet + \alf}{2} \rpar \pi \rbk}{\sin \lbk \lpar \mu - \frac{\bet +
\alf}{2} \rpar \frac{\pi}{N} \rbk} \cos \lbk \frac{2 \pi \nu}{N} (\bet - \alf) \rbk  \qquad \mbox{with} \;\; \alf \neq \bet ,} \\
\lb{eq19s3}
\eqalign{
\fl \qquad \bigl( \hat{\mathscr{V}}_{\alf,\bet} \bigr)(\mu,\nu) &= 2 \delta_{\mu,\veps}^{[N]} \sin \lbk \frac{2 \pi \nu}{N} 
(\bet - \alf) \rbk \qquad \mbox{if} \;\; \alf + \bet = 2 \veps \;\; \mbox{and} \;\; \veps \in \mathbb{N}^{\ast} , \cr
\fl \qquad &= \frac{2}{N} \frac{\sin \lbk \lpar \mu - \frac{\bet + \alf}{2} \rpar \pi \rbk}{\sin \lbk \lpar \mu - \frac{\bet +
\alf}{2} \rpar \frac{\pi}{N} \rbk} \sin \lbk \frac{2 \pi \nu}{N} (\bet - \alf) \rbk  \qquad \mbox{with} \;\; \alf \neq \bet ,} \\
\lb{eq20s3}
\fl \qquad \bigl( \hat{\mathscr{W}}_{\gam} \bigr)(\mu,\nu) = \sqrt{\frac{2}{(\gam+1)(\gam+2)}} \lbk \sum_{\sigma = 0}^{\gam}
\delta_{\mu,\sigma}^{[N]} - (\gam + 1) \delta_{\mu,\gam+1}^{[N]} \rbk .
\err
Therefore, given a physical system with $\hat{\rho}$ described by $\mathrm{SU(N)}$, the terms $\{ \lg \hat{g}_{i} \rg \}_{i = 1,
\ldots,N^{2}-1}$ can be promptly determined in this case, and consequently, the discrete Wigner function \eref{eq17s3} properly
established. The time evolution of $\mathit{W}(\mu,\nu;t)$ is introduced via $\hat{\rho}(t)$.

To illustrate our results let us consider the group $\mathrm{SU(2)}$ once again. So, the discrete Wigner function for two-level
systems \cite{Fano} is given in this context by
\bd
\fl \qquad \mathit{W}(\mu,\nu) = \half \lbk 1 + (-1)^{\nu} P_{x} + (-1)^{\mu+\nu+1} P_{y} + (-1)^{\mu} P_{z} \rbk \qquad 
(0 \leq \mu,\nu \leq 1) , 
\ed
where $P_{i} = \Tr [ \hat{\rho} \hat{\sigma}_{i} ] \in [-1,1]$ for $i=x,y,z$ represents the components of the polarization vector
$\mathbf{P}$ with $P_{x}^{2} + P_{y}^{2} + P_{z}^{2} \leq 1$ (the saturation occurs for pure states). Moreover, this function can 
be measured since the components of $\mathbf{P}$ are experimentally obtained \cite{Paris-book,Teo-book}. However, it is worth to
emphasize that different $\mathrm{SU(2)}$ Wigner function approaches using continuous variables have emerged from the discussion
on atomic coherent states in the recent past \cite{Ag1981,CS1986,DAS1994,Ag1998,RGS2001,ST2011}. Notwithstanding their inherent
technical difficulties, the approaches were applied with broad success on a collection of two-level systems \cite{DAS1994,Ag1998}
or even in the tomographic reconstruction of a spin-squeezed state of a Bose-Einstein condensate \cite{ST2011}. Finally, let us 
say some few words about the $\mathrm{SU(2)}$ Wigner function here exhibited: it corresponds to a genuinely discrete 
quasiprobability distribution function and obeys the criterion `easy-to-handle' -- see also Ref. \cite{KRG2017}.

With respect to the discrete $\mathrm{SU(N)}$ Wigner function \eref{eq17s3}, its elegance and simplicity hide possible 
difficulties inherent to the calculation of generators for high values of $N$, as well as the determination of the elements
$\lg \hat{g}_{i} \rg$ and $\lpar \hat{g}_{i} \rpar \! (\mu,\nu)$ for all $i=1,\ldots,N^{2}-1$. This apparent disadvantage can be
circumvented through the symmetries and dynamical characteristics of the physical system under investigation \cite{MSG2009}.
Tilma and coworkers \cite{TN2012} showed, a few years ago, that generalized $\mathrm{SU(N)}$-symmetric coherent states can be used 
to construct the continuous quasiprobability distribution functions (namely, the Wigner, Husimi, and Glauber-Sudarshan functions
\cite{Orszag-book}); subsequently, the authors proposed an alternative framework for computing Wigner functions which describe
physical systems with arbitrary dimensions \cite{TESMN}. Therefore, \Eref{eq17s3} can be interpreted as a discrete version of
their results.

\section{Application: the group $\mathbf{SU(3)}$}
\lb{s4}

Up to now the first example treated throughout the text was the group $\mathrm{SU(2)}$, namely, a single two-level quantum system
where a general qubit state can be promptly represented by means of its discrete Wigner function. The next natural step is to
obtain the discrete Wigner function for three-level quantum systems related to the group $\mathrm{SU(3)}$ and mainly described by
\be
\lb{eq21s4}
\ro = \left( \begin{array}{ccc}
\rho_{11}        & \rho_{12}        & \rho_{13} \\
\rho_{12}^{\ast} & \rho_{22}        & \rho_{23} \\
\rho_{13}^{\ast} & \rho_{23}^{\ast} & \rho_{33} \\
\end{array} \right) \in \mathscr{L}_{+,1}(\mathcal{H}_{3}) \, .
\ee
For such a task, let us initially consider the results established for the $\mathrm{SU(3)}$ generators in \ref{apb}, as well as
their respective mapped expressions $\{ \bigl( \hat{\lam}_{i} \bigr)(\mu,\nu) \}_{i=1,\ldots,8}$ in the finite-dimensional
discrete phase space. These results allow, in particular, to determine compact expressions for the components $\bigl( \lg 
\hat{\lam}_{1} \rg,\ldots,\lg \hat{\lam}_{8} \rg \bigr)$ of the Bloch vector $\mathbf{g}$,
\brr
\lb{eq22s4}
\eqalign{\lg \hat{\lam}_{1} \rg = 2 \, \mathtt{Re} (\rho_{12}) , \; \lg \hat{\lam}_{2} \rg = - 2 \, \mathtt{Im} (\rho_{12}) , \;
\lg \hat{\lam}_{3} \rg = \rho_{11} - \rho_{22} , \\
\lg \hat{\lam}_{4} \rg = 2 \, \mathtt{Re} (\rho_{13}) , \; \lg \hat{\lam}_{5} \rg = - 2 \, \mathtt{Im} (\rho_{13}) , \; 
\lg \hat{\lam}_{6} \rg = 2 \, \mathtt{Re} (\rho_{23}) , \\
\lg \hat{\lam}_{7} \rg = - 2 \, \mathtt{Im} (\rho_{23}) , \; \lg \hat{\lam}_{8} \rg = \frac{\sqrt{3}}{3} ( \rho_{11} + 
\rho_{22} - 2 \rho_{33} ) .}
\err
In this way, the discrete $\mathrm{SU(3)}$ Wigner function is properly established with the help of \Eref{eq17s3} for $N=3$,
\brr
\lb{eq23s4}
\fl \mathit{W}(\mu,\nu) = \frac{1}{3} + \frac{1}{3} \bigl( 2 \delta_{\mu,0}^{[3]} - \delta_{\mu,1}^{[3]} - 
\delta_{\mu,2}^{[3]} \bigr) \rho_{11} - \frac{1}{3} \bigl( \delta_{\mu,0}^{[3]} - 2 \delta_{\mu,1}^{[3]} + \delta_{\mu,2}^{[3]}
\bigr) \rho_{22} \nn \\
\fl \qquad \qquad \;\; - \frac{1}{3} \bigl( \delta_{\mu,0}^{[3]} + \delta_{\mu,1}^{[3]} - 2 \delta_{\mu,2}^{[3]} \bigr) 
\rho_{33} + 2 \delta_{\mu,1}^{[3]} \lbk \cos \lpar \frac{4 \pi \nu}{3} \rpar \mathtt{Re} (\rho_{13}) - \sin \lpar 
\frac{4 \pi \nu}{3} \rpar \mathtt{Im} (\rho_{13}) \rbk \nn \\
\fl \qquad \qquad \;\; + \frac{2}{3} \frac{\sin \lbk \lpar \mu - \half \rpar \pi \rbk}{\sin \lbk \lpar \mu - \half \rpar
\frac{\pi}{3} \rbk} \lbk \cos \lpar \frac{2 \pi \nu}{3} \rpar \mathtt{Re} (\rho_{12}) - \sin \lpar \frac{2 \pi \nu}{3} \rpar
\mathtt{Im} (\rho_{12}) \rbk \nn \\
\fl \qquad \qquad \;\; + \frac{2}{3} \frac{\sin \lbk \lpar \mu - \frac{3}{2} \rpar \pi \rbk}{\sin \lbk \lpar \mu - \frac{3}{2}
\rpar \frac{\pi}{3} \rbk} \lbk \cos \lpar \frac{2 \pi \nu}{3} \rpar \mathtt{Re} (\rho_{23}) - \sin \lpar \frac{2 \pi \nu}{3} \rpar
\mathtt{Im} (\rho_{23}) \rbk .
\err
This expression is completely general and can be particularly applied to the description of a qutrit
\cite{Men2006,Kur2011,GSSS2016}, and its three-dimensional visualization represents a complementary result to that discussed 
in Ref. \cite{KKLM2016} -- see \Tref{tab1} for $0 \leq \mu,\nu \leq 2$.
\begin{table}[!t]
\centering
\caption{\lb{tab1} Possible values of the discrete Wigner functions \eref{eq23s4} and \eref{eq25s4}.}
\footnotesize
\begin{tabular}{@{}llll}
\br
$\mu$ & $\nu$ & \Eref{eq23s4} & \Eref{eq25s4} \\
\mr
0 & 0 & $\rho_{11} + \frac{2}{3} \mathtt{Re} ( 2 \rho_{12} + \rho_{23} )$ & $\frac{1}{3} + \frac{2}{3} ( 2 \wp_{1} + \wp_{3} )$ \\
0 & 1 & $\rho_{11} - \frac{1}{3} \mathtt{Re} ( 2 \rho_{12} + \rho_{23} ) - \frac{\sqrt{3}}{3} \mathtt{Im} ( 2 \rho_{12} + 
\rho_{23} )$ & $\frac{1}{3} - \frac{1}{3} ( 2 \wp_{1} + \wp_{3} )$ \\
0 & 2 & $\rho_{11} - \frac{1}{3} \mathtt{Re} ( 2 \rho_{12} + \rho_{23} ) + \frac{\sqrt{3}}{3} \mathtt{Im} ( 2 \rho_{12} + 
\rho_{23} )$ & $\frac{1}{3} - \frac{1}{3} ( 2 \wp_{1} + \wp_{3} )$ \\
1 & 0 & $\rho_{22} + \frac{4}{3} \mathtt{Re} \lpar \rho_{12} + \frac{3}{2} \rho_{13} + \rho_{23} \rpar$ & $\frac{1}{3} + 
\frac{4}{3} \lpar \wp_{1} + \frac{3}{2} \wp_{2} + \wp_{3} \rpar$ \\
1 & 1 & $\rho_{22} - \frac{2}{3} \mathtt{Re} \lpar \rho_{12} + \frac{3}{2} \rho_{13} + \rho_{23} \rpar - \frac{\sqrt{3}}{3}
\mathtt{Im} ( 2 \rho_{12} - 3 \rho_{13} + 2 \rho_{23} )$ & $\frac{1}{3} - \frac{2}{3} \lpar \wp_{1} + \frac{3}{2} \wp_{2} +
\wp_{3} \rpar$ \\
1 & 2 & $\rho_{22} - \frac{2}{3} \mathtt{Re} \lpar \rho_{12} + \frac{3}{2} \rho_{13} + \rho_{23} \rpar + \frac{\sqrt{3}}{3}
\mathtt{Im} ( 2 \rho_{12} - 3 \rho_{13} + 2 \rho_{23} )$ & $\frac{1}{3} - \frac{2}{3} \lpar \wp_{1} + \frac{3}{2} \wp_{2} +
\wp_{3} \rpar$ \\
2 & 0 & $\rho_{33} - \frac{2}{3} \mathtt{Re} ( \rho_{12} - 2 \rho_{23} )$ & $\frac{1}{3} - \frac{2}{3} (\wp_{1} - 2 \wp_{3})$ \\
2 & 1 & $\rho_{33} + \frac{1}{3} \mathtt{Re} ( \rho_{12} - 2 \rho_{23} ) + \frac{\sqrt{3}}{3} \mathtt{Im} ( \rho_{12} - 
2 \rho_{23} )$ & $\frac{1}{3} + \frac{1}{3} ( \wp_{1} - 2 \wp_{3})$ \\
2 & 2 & $\rho_{33} + \frac{1}{3} \mathtt{Re} ( \rho_{12} - 2 \rho_{23} ) - \frac{\sqrt{3}}{3} \mathtt{Im} ( \rho_{12} - 
2 \rho_{23} )$ & $\frac{1}{3} + \frac{1}{3} ( \wp_{1} - 2 \wp_{3})$ \\
\br
\end{tabular}
\end{table}

\begin{figure}[t]
\begin{minipage}[b]{0.3\linewidth}
\includegraphics[width=\linewidth]{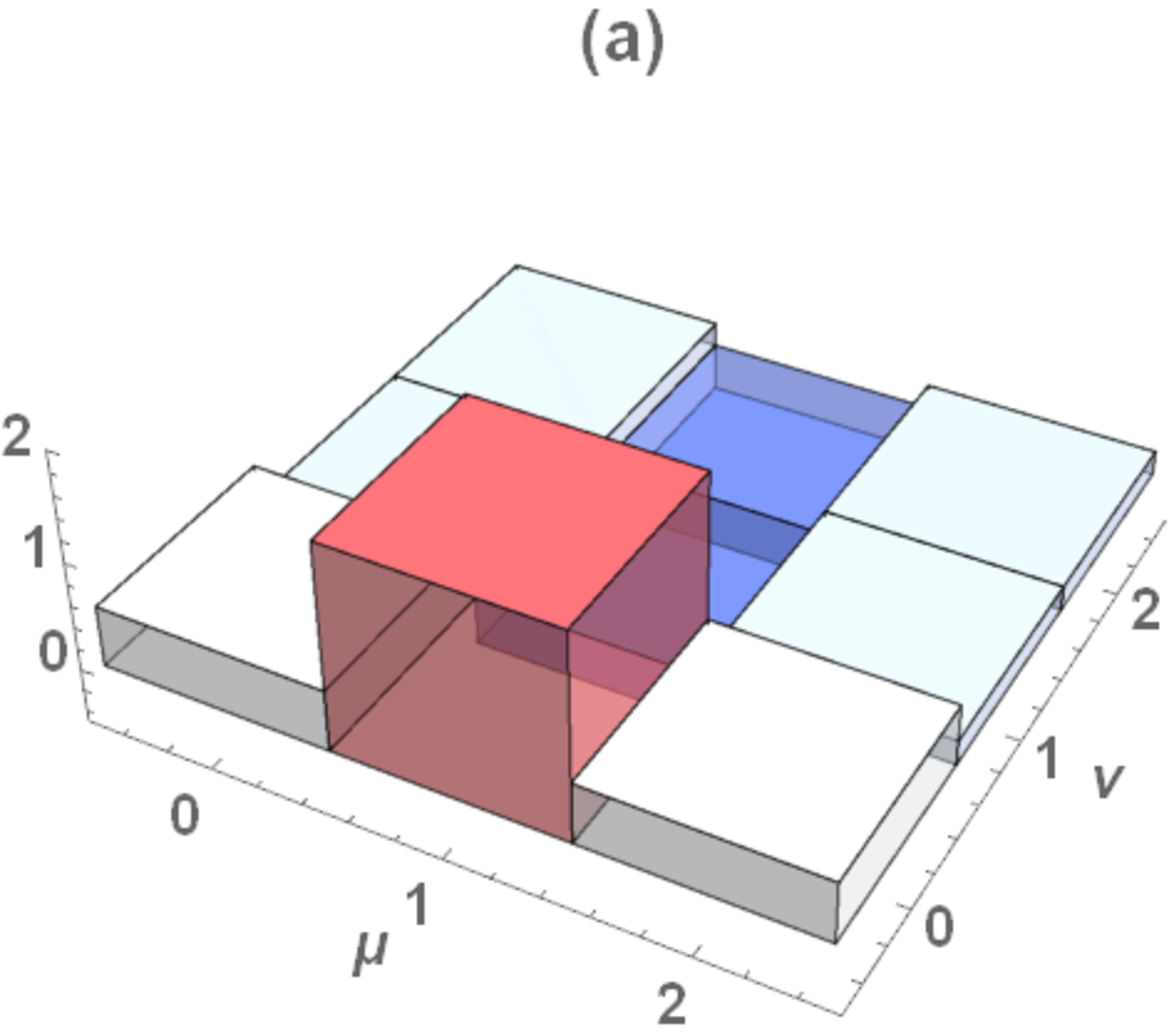}
\end{minipage} \hfill
\begin{minipage}[b]{0.3\linewidth}
\includegraphics[width=\linewidth]{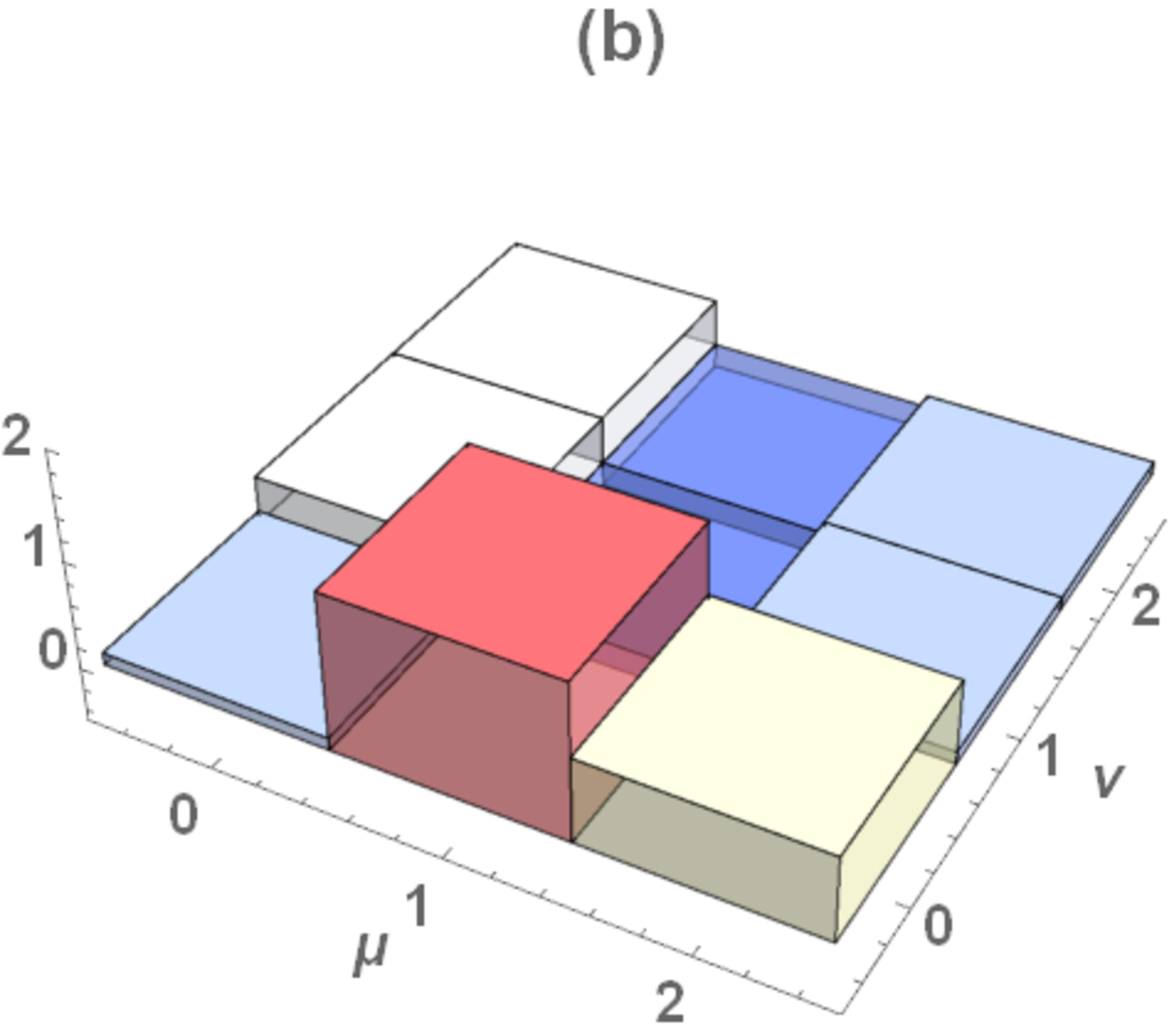}
\end{minipage} \hfill
\begin{minipage}[b]{0.3\linewidth}
\includegraphics[width=\linewidth]{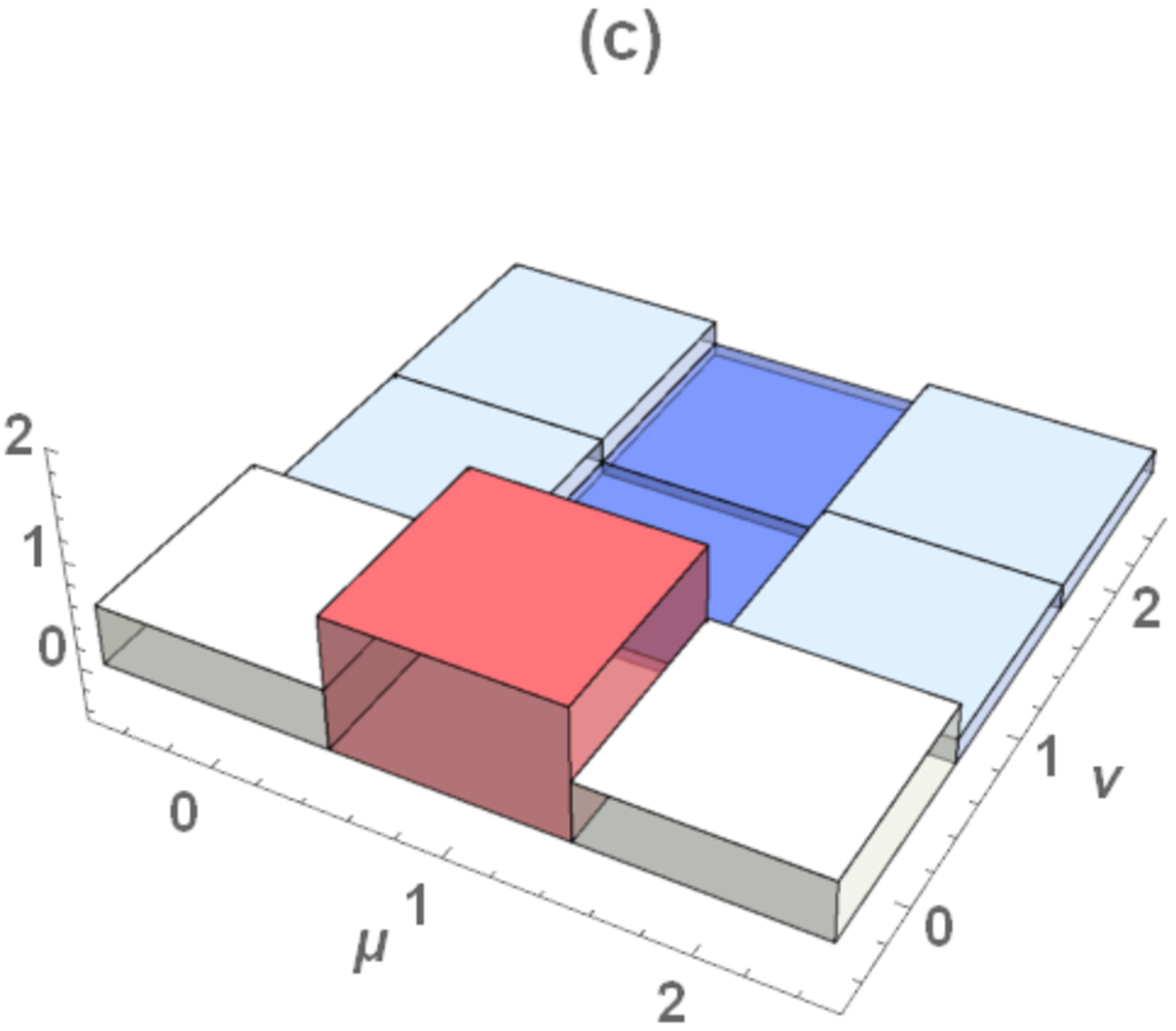}
\end{minipage} \hfill
\begin{minipage}[b]{0.3\linewidth}
\includegraphics[width=\linewidth]{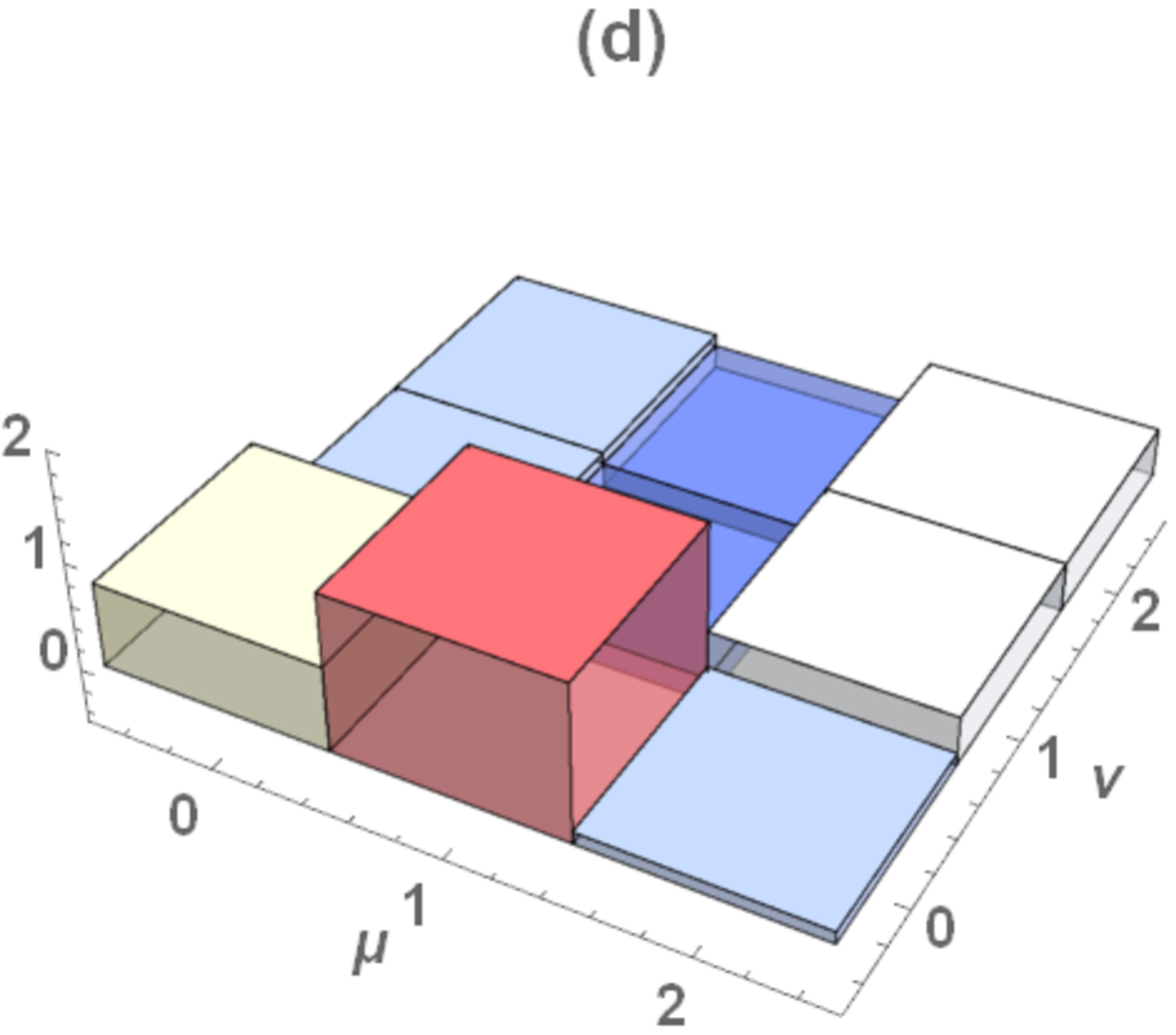}
\end{minipage} \hfill
\begin{minipage}[b]{0.3\linewidth}
\includegraphics[width=\linewidth]{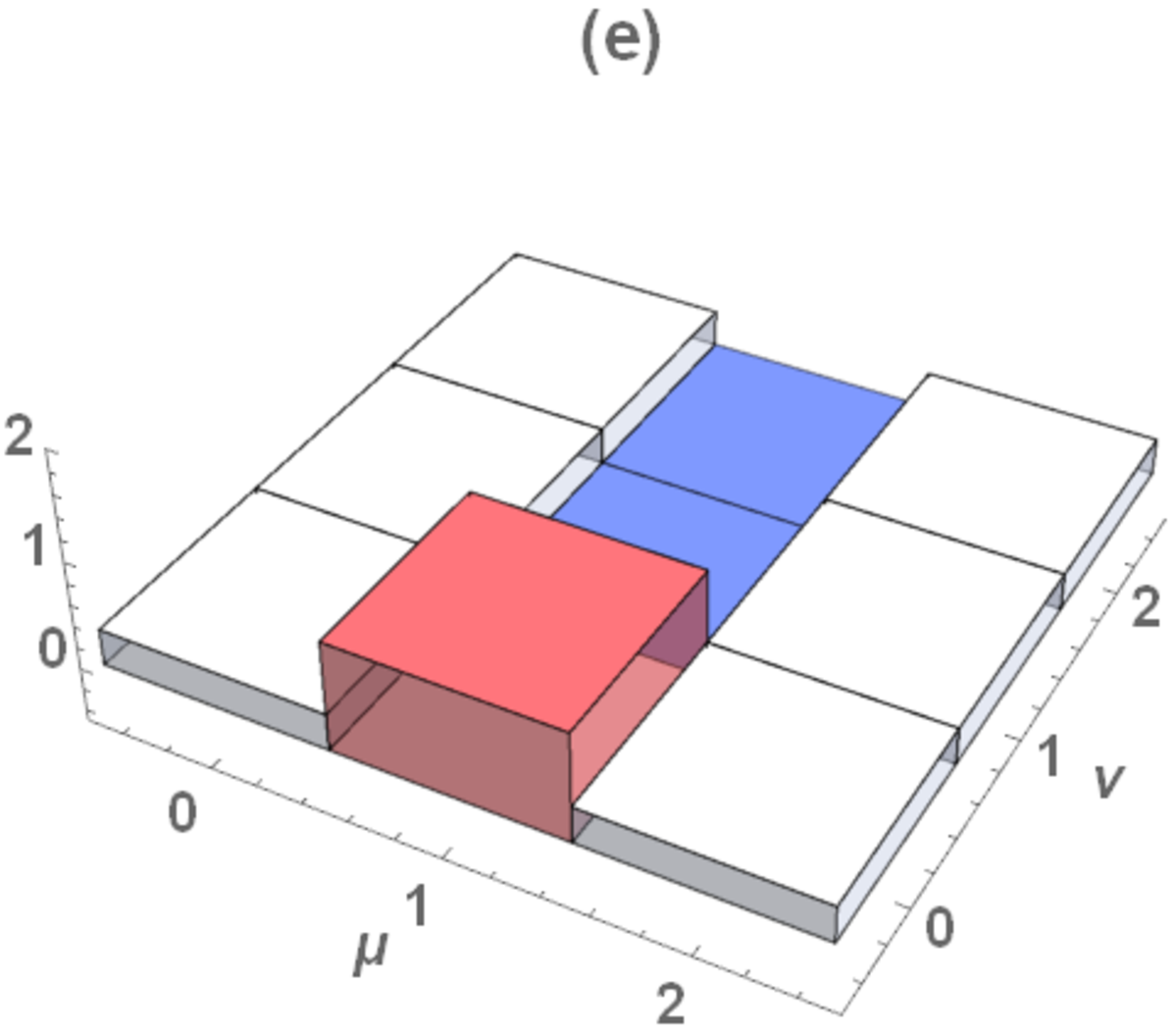}
\end{minipage} \hfill
\begin{minipage}[b]{0.3\linewidth}
\includegraphics[width=\linewidth]{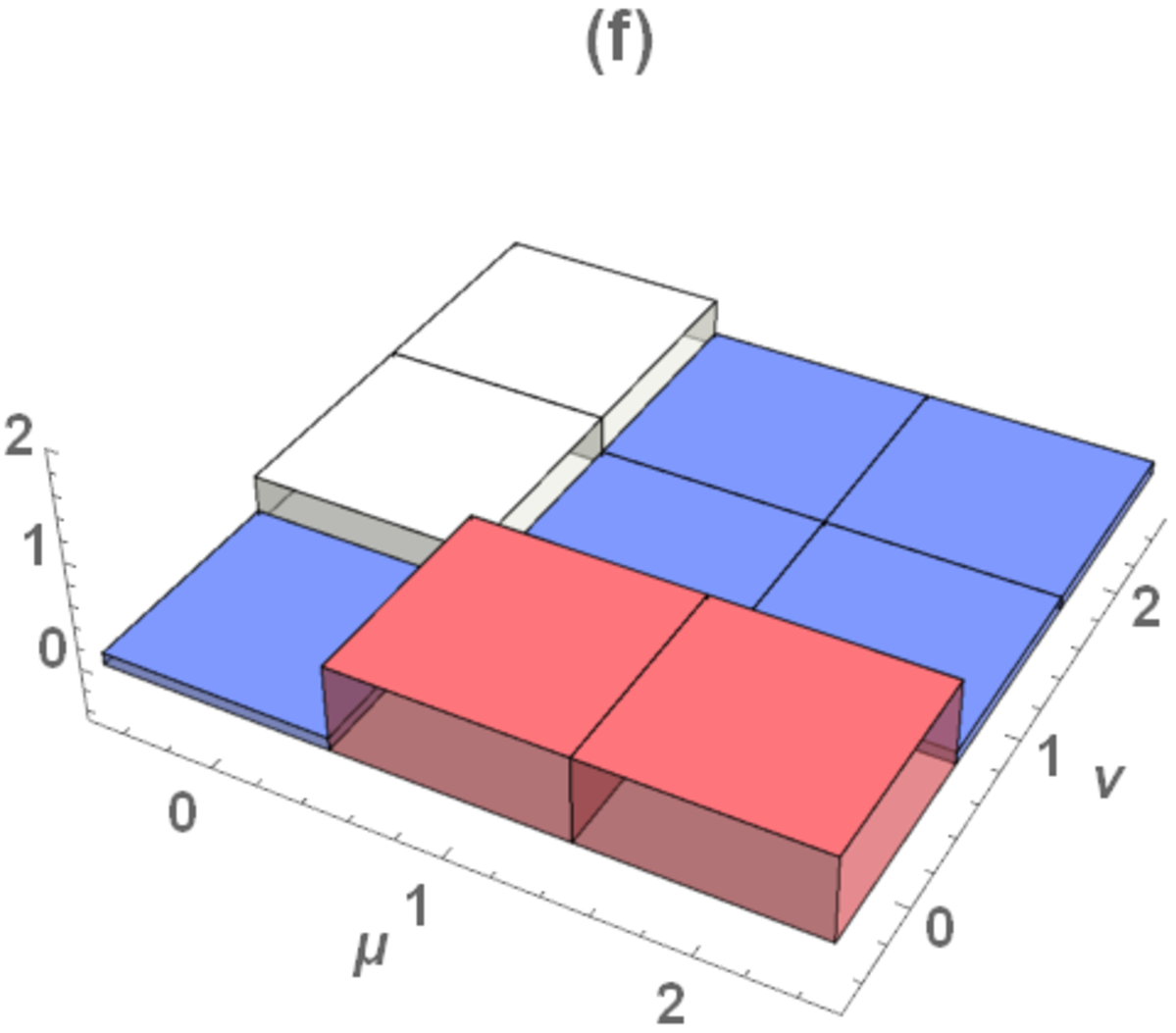}
\end{minipage}
\caption{Three-dimensional plots of the discrete Wigner function $\mathit{W}(\mu,\nu)$ described by \Eref{eq25s4} versus 
$(\mu,\nu) \in [0,2]$ with $N=3$ fixed, and different transition rates: (a) $\wp_{1}=\wp_{2}=\wp_{3}=\frac{1}{3}$, 
(b) $\wp_{1}=0$ and $\wp_{2}=\wp_{3}=\frac{1}{3}$, (c) $\wp_{1}=\wp_{3}=\frac{1}{3}$ and $\wp_{2}=0$, (d) $\wp_{1}=\wp_{2}=
\frac{1}{3}$ and $\wp_{3}=0$, (e) $\wp_{1}=\wp_{3}=0$ with $\wp_{2}=\frac{1}{3}$; and finally, (f) $\wp_{1}=\wp_{2}=0$ and
$\wp_{3} = \frac{1}{3}$. In such a case, (a) represents a pure state with $\mathit{W}(1,1) = \mathit{W}(1,2) \approx - 0,44$ and
$\mathit{W}(1,0) \approx 1,89$ reflecting its respective minimum and maximum values. Moreover, (e) and (f) depict situations where
the discrete Wigner function is positive.}
\end{figure}
To illustrate this result, we consider a specific three-level system described by
\be
\lb{eq24s4}
\ro = \left( \begin{array}{ccc}
\frac{1}{3} & \wp_{1}     & \wp_{2} \\
\wp_{1}     & \frac{1}{3} & \wp_{3} \\
\wp_{2}     & \wp_{3}     & \frac{1}{3} \\
\end{array} \right) \qquad \lpar \wp_{1},\wp_{2},\wp_{3} \in \mathbb{R}_{+} \rpar ,
\ee
where $\wp_{1}$, $\wp_{2}$, and $\wp_{3}$ are associated with the respective transition rates between the states 
$0 \rightleftharpoons 1$, $0 \rightleftharpoons 2$, and $1 \rightleftharpoons 2$, these states being equally populated in the
ratio of $\frac{1}{3}$. In this toy model, the condition $\wp_{1}^{2} + \wp_{2}^{2} + \wp_{3}^{2} \leq \frac{1}{3}$ asserts that
$\rho_{\ell} \in \mathbb{R}_{+}$ for $\ell=1,2,3$ (namely, the eigenvalues assume real and positive values), the saturation being
reached in this case only for pure states. Following, the discrete Wigner function \eref{eq23s4} can be expressed in such a case
as follows:
\brr
\lb{eq25s4}
\fl \qquad \quad \mathit{W}(\mu,\nu) =& \frac{1}{3} + \frac{2}{3} \frac{\sin \lbk \lpar \mu - \half \rpar \pi \rbk}{\sin \lbk 
\lpar \mu - \half \rpar \frac{\pi}{3} \rbk} \cos \lpar \frac{2 \pi \nu}{3} \rpar \wp_{1} + 2 \delta_{\mu,1}^{[3]} \cos \lpar 
\frac{4 \pi \nu}{3} \rpar \wp_{2} \nn \\
\fl \qquad \quad &+ \frac{2}{3} \frac{\sin \lbk \lpar \mu - \frac{3}{2} \rpar \pi \rbk}{\sin \lbk \lpar \mu - \frac{3}{2} \rpar 
\frac{\pi}{3} \rbk} \cos \lpar \frac{2 \pi \nu}{3} \rpar \wp_{3} \, .
\err
Figure 1 shows the 3D plots of this expression as a function of $(\mu,\nu)$ for different values of $\wp_{1}$, $\wp_{2}$, and 
$\wp_{3}$. For instance, the negative values appearing in (a-d) at the points $(1,1)$ and $(1,2)$ represent a quantum signature of
the nonclassical effects associated with the particular three-level system under investigation (in these cases, at least two
transitions are allowed). However, (e-f) exhibit only positive values related to the transitions $0 \rightleftharpoons 2$ or 
$1 \rightleftharpoons 2$ -- see \Tref{tab1}. Note that this specific toy model can be promptly generalized in order to incorporate
more realistic effects such as different population rates and complex transition rates.

Recently, Martins and coworkers \cite{MKG2019} exhibited a set of results on the $\star$-product for $\mathrm{SU(3)}$ Wigner 
functions over $\mathrm{SU(3)/U(2)}$. In particular, these functions were defined in a symplectic manifold which corresponds to a
classical phase space that is, by its turn, related to a specific set of orbit-type coherent states. It is worth stressing that
\Eref{eq23s4} describes the $\mathrm{SU(3)}$ Wigner function defined upon a finite-dimensional phase space labelled by genuine
discrete variables associated with spin representations, which differs from that aforementioned result.

\section{Concluding remarks}
\lb{s5}

In this paper, we have established a self-contained theoretical framework for the discrete Wigner function associated with a wide
class of arbitrary quantum systems characterized by a finite space of states. Indeed the connection between $\mathrm{SU(N)}$
generators and Schwinger unitary operators (and vice versa) via the $\mathit{mod(N)}$-invariant unitary operator basis paves
the way to introduce a finite-dimensional phase space which is genuinely discrete. Thus, the discrete Wigner function obtained
from this guideline is completely general since it allows, within other possibilities, to describe arbitrary spin systems and 
also to provide a theoretical background for nonrelativistic studies on particle physics models. Next, let us focus our attention 
on effective gains and future perspectives derived from this manuscript which deserve to be properly discussed.

\begin{itemize}
\item The mathematical framework exposed here can be promptly generalized in order to include the extended Cahill-Glauber
formalism for finite-dimensional spaces \cite{RMG2005,MRG2005}. For this task, it is enough to substitute the 
$\mathit{mod(N)}$-invariant operator basis $\hat{G}(\mu,\nu)$ by its extended version
\bd
\fl \qquad \qquad \hat{T}^{(s)}(\mu,\nu) = \frac{1}{\sqrt{N}} \sum_{\eta,\xi=-\ell}^{\ell} \om^{-(\mu \eta + \nu \xi)} 
\om^{\half N \Phi(\eta,\xi;N)} \lbk \mathcal{K}(\eta,\xi) \rbk^{-s} \hat{S}_{\mathrm{S}}(\eta,\xi) ,
\ed
where the additional term $\mathcal{K}(\eta,\xi)$ is expressed through a nontrivial sum of products of Jacobi theta functions for
integer arguments with $|s| \leq 1$ \cite{RMG2005}. So, \Eref{eq17s3} assumes the general form
\be
\lb{eq26s4}
\fl \qquad F^{(s)}(\mu,\nu) \col \Tr [ \hat{T}^{(s)}(\mu,\nu) \ro ] = \frac{1}{N} + \half \sum_{i=1}^{N^{2}-1} \lg \hat{g}_{i} \rg
\underbrace{\bigl( \hat{g}_{i} \bigr)^{(s)}(\mu,\nu)}_{\Tr [ \hat{T}^{(s)}(\mu,\nu) \hat{g}_{i} ]} \, ,
\ee
such that for $s=-1,0,+1$ the parametrized function $F^{(s)}(\mu,\nu)$ recovers the discrete Husimi, Wigner, and Glauber-Sudarshan
functions, respectively. 

\item A first application of this quantum-algebraic framework is associated with the study of bipartite systems. For example, 
let $\mathcal{H}_{N_{1}}$ and $\mathcal{H}_{N_{2}}$ denote both the Hilbert spaces of the parts $1$ and $2$, as well as 
$\mathcal{H}_{N} = \mathcal{H}_{N_{1}} \otimes \mathcal{H}_{N_{2}}$ characterize the Hilbert space of the total system for
$N = N_{1} N_{2}$. In addition, let $\{ \hat{g}_{1,i} \}_{i=1,\ldots,N_{1}^{2}-1}$ and $\{ \hat{g}_{2,j} \}_{j=1,\ldots,
N_{2}^{2}-1}$ be the respective generators of the groups $\mathrm{SU(N_{1})}$ and $\mathrm{SU(N_{2})}$ used to describe each part,
with $\mathrm{SU(N)} = \mathrm{SU(N_{1})} \otimes \mathrm{SU(N_{2})}$ being responsible for bipartite system. From the kinematical
point of view, there are two distinct approaches to deal with such a system via discrete Wigner functions: the first one 
corresponds to write the density matrix $\ro$ of the total system in terms of the aforementioned generators $\hat{g}_{1,i}$ and 
$\hat{g}_{2,j}$ related to the parts $1$ and $2$ through the techniques developed in Refs. \cite{Fano1957,BMMP2012}, and then
obtain the corresponding discrete Wigner function; the second one consists of using \Eref{eq17s3} as a guideline, where now the
generators $\{ \hat{g}_{k} \}_{k=1,\ldots,N^{2}-1}$ of the total system entered the scene. Following, two-qubits X-states
\cite{MMG2014} and qubit-qutrit systems \cite{MMH2017} represent two typical examples where these approaches can be applied to the
study on entanglement effects of bipartite systems, and also in the study of hybrid systems \cite{RCRT}.
\end{itemize}

To conclude, let us discuss some pertinent points associated with more general spin systems and multipartite quantum states. It is
worth of mention that $\hat{G}(\mu,\nu)$ was already used in the description of spin squeezing effects in straight connection with
entaglement properties of the modified Lipkin-Meshkov-Glick model via discrete Wigner function for high values of $N$
\cite{MGD2013} -- see Ref. \cite{MSG2009} for a discussion on spin-tunneling processes involving an original version of the
aforementioned model, where now the discrete Husimi function has been extensively used. Moreover, the active and fruitful field
of research on N-qubit X-states \cite{MRGM2015} represents, nowadays, an interesting scenario of possible applications for
discrete Wigner function where the study on maximally genuine multipartite entangled mixed states will take place \cite{SC2019}.

\appendix
\section{The parity operator}
\lb{apa}

The discussion on the equivalence between large $N$ limits of quantum theories and classical limits certainly reveals a heated
historical debate that has dragged on for many decades -- for instance, see Refs. \cite{Yaffe1982,Yaffe1983,APD2017}. In this
appendix we will show, in particular, how the continuum limit of a genuinely discrete quantum approach takes place, emphasizing, 
by its turn, the continuum limit $N \rightarrow \infty$ of $\hat{G}(0,0)$ in connection with the continuous parity operator. 
Furthermore, we will also present an alternative form of the $\mathit{mod(N)}$-invariant unitary operator basis written in terms 
of the discrete parity operator.

\begin{definition}
Let $\hat{\mathfrak{F}}$ denote the finite-dimensional discrete Fourier operator defined in terms of the set $\{ | u_{\bet} \rg,
| v_{\bet} \rg \}_{\bet=0,\ldots,N-1}$ of orthonormal eigenvectors,
\be
\lb{eqa1}
\fl \qquad \hat{\mathfrak{F}} \col \sum_{\bet=0}^{N-1} | v_{\bet} \rg \lg u_{\bet} | = \frac{1}{\sqrt{N}} 
\sum_{\bet,\bet^{\prime}=0}^{N-1} \om^{\bet \bet^{\prime}} | u_{\bet^{\prime}} \rg \lg u_{\bet} | \Rightarrow \hat{\mathfrak{F}}
\hat{\mathfrak{F}}^{\dagger} = \hat{\mathfrak{F}}^{\dagger} \hat{\mathfrak{F}} = \hat{\mathds{I}} . 
\ee
The additional properties \cite{Me1987,CD2013}
\be
\lb{eqa2}
\fl \qquad \hat{\mathfrak{F}}^{2} = \sum_{\bet=0}^{N-1} | u_{-\bet} \rg \lg u_{\bet} | , \quad \hat{\mathfrak{F}}^{3} = 
\sum_{\bet=0}^{N-1} | v_{-\bet} \rg \lg u_{\bet} | , \;\; \mbox{and} \quad \hat{\mathfrak{F}}^{4} = \hat{\mathds{I}} ,
\ee
lead us to find out that $\hat{\mathfrak{F}}$ is a periodic operator with $4$-period. Furthermore, the property associated with
$\hat{\mathfrak{F}}^{2}$ also permits to establish the discrete parity operator $\hat{\mathsf{P}}$ by means of the relation 
$\hat{\mathsf{P}} \col \hat{\mathfrak{F}}^{2}$.
\end{definition}

\begin{remark}
Now, let us clarify an important point related to the $\mathit{mod(N)}$-invariant unitary operator basis \eref{eq3s2}. Initially,
we should observe that
\bd
\lbk \sqrt{N} \hat{S}_{\mathrm{S}}(\nu,-\mu) \rbk \hat{S}_{\mathrm{S}}(\eta,\xi) \lbk \sqrt{N} \hat{S}_{\mathrm{S}}^{\dagger}
(\nu,-\mu) \rbk = \om^{-( \mu \eta + \nu \xi )} \hat{S}_{\mathrm{S}}(\eta,\xi)
\ed
depicts a similarity transformation, this particular result being responsible for rewriting $\hat{G}(\mu,\nu)$ as follows:
\be
\lb{eqa3}
\hat{G}(\mu,\nu) = \lbk \sqrt{N} \hat{S}_{\mathrm{S}}(\nu,-\mu) \rbk \hat{G}(0,0) \lbk \sqrt{N} \hat{S}_{\mathrm{S}}^{\dagger}
(\nu,-\mu) \rbk ,
\ee
where
\bd
\fl \qquad \; \lbk \sqrt{N} \hat{S}_{\mathrm{S}}(\nu,-\mu) \rbk \!\! \lbk \sqrt{N} \hat{S}_{\mathrm{S}}^{\dagger}(\nu,-\mu) \rbk =
\lbk \sqrt{N} \hat{S}_{\mathrm{S}}^{\dagger}(\nu,-\mu) \rbk \!\! \lbk \sqrt{N} \hat{S}_{\mathrm{S}}(\nu,-\mu) \rbk = 
\hat{\mathds{I}} .
\ed
Therefore, \Eref{eqa3} not only represents an unitary transformation of $\hat{G}(0,0)$, but also establishes a displacement of a
given initial point $(0,0)$ to the final point $(\mu,\nu)$ on the associated operator space -- for a discussion on construction
and properties of the discrete coherent states, see Ref. \cite{GM1996}. Furthermore, note that $\hat{G}(0,0)$ apparently does not
exhibit any connection with the previously defined discrete parity operator.
\end{remark}

An interesting question then emerges from our considerations on Schwinger unitary operators and $\mathit{mod(N)}$-invariant
unitary operator basis $\hat{G}(\mu,\nu)$: ``Can $\hat{G}(0,0)$ describe the parity operator in the continuum limit $N \rightarrow
\infty$"? To answer this particular question, let us initially adopt the mathematical prescription for the continuum limit
established in Refs. \cite{RG2000,RG2002} through the following steps:
\begin{itemize}
\item Let $\hat{Q}$ and $\hat{P}$ denote, respectively, the discrete coordinate and momentum operators defined in a 
$N$-dimensional state vector space, which obey the eigenvalue equations
\bd
\hat{Q} | \mathfrak{q}_{\alf} \rg = \mathfrak{q}_{\alf} | \mathfrak{q}_{\alf} \rg \quad \mbox{and} \quad 
\hat{P} | \mathfrak{p}_{\bet} \rg = \mathfrak{p}_{\bet} | \mathfrak{p}_{\bet} \rg ,
\ed
where $\mathfrak{q}_{\alf} \col \varepsilon q_{0} \alf$ and $\mathfrak{p}_{\bet} \col \varepsilon p_{0} \bet$ represent two
distinct sets of discrete eigenvalues labeled by $\alf,\bet \in [-\ell,\ell]$ for $\ell = \frac{N-1}{2}$ and $N$ odd, with 
$\varepsilon = \sqrt{\frac{2 \pi}{N}}$ fixed. For simplicity, the distances between sucessive eigenvalues of $\hat{Q}$ and 
$\hat{P}$ are maintained constants, that is, $D_{\mathfrak{q}} = \varepsilon q_{0}$ and $D_{\mathfrak{p}} = \varepsilon p_{0}$.
Besides, the real parameters $q_{0}$ (coordinate unity) and $p_{0}$ (momentum unity) satisfy the relation $q_{0} p_{0} = \hbar$;
consequently, the discrete eigenvalues can be written as $\mathfrak{q}_{\alf} = D_{\mathfrak{q}} \alf$ and $\mathfrak{p}_{\bet} =
D_{\mathfrak{p}} \bet$.

\item The connection between $(\hat{Q},\hat{P})$ and $(\hat{U},\hat{V})$ follows the mathematical prescription
\bd
\hat{U} = \exp \lpar \frac{\im}{\hbar} D_{\mathfrak{p}} \hat{Q} \rpar \quad \mbox{and} \quad \hat{V} = \exp \lpar 
\frac{\im}{\hbar} D_{\mathfrak{q}} \hat{P} \rpar ,
\ed
which implies that $\om^{\half \eta \xi} \hat{U}^{\eta} \hat{V}^{\xi}$ present in $\hat{G}(0,0)$ leads to the expression
\bd
\om^{\half \eta \xi} \hat{U}^{\eta} \hat{V}^{\xi} = \exp \lpar - \frac{\im}{2 \hbar} \mathfrak{p}_{\eta} \mathfrak{q}_{\xi} \rpar
\exp \lpar \frac{\im}{\hbar} \mathfrak{p}_{\eta} \hat{Q} \rpar \exp \lpar - \frac{\im}{\hbar} \mathfrak{q}_{\xi} \hat{P} \rpar
\ed
with $\mathfrak{p}_{\eta} = D_{\mathfrak{p}} \eta$ and $\mathfrak{q}_{\xi} = - D_{\mathfrak{q}} \xi$. Therefore, $\hat{G}(0,0)$
assumes the form
\bd
\fl \quad \hat{G}(0,0) = \sum_{\mathfrak{p}_{\eta} = - R_{\mathfrak{p}}}^{R_{\mathfrak{p}}} \sum_{\mathfrak{q}_{\xi} = - 
R_{\mathfrak{q}}}^{R_{\mathfrak{q}}} \frac{D_{\mathfrak{p}} D_{\mathfrak{q}}}{2 \pi \hbar} \exp \lpar - \frac{\im}{2 \hbar}
\mathfrak{p}_{\eta} \mathfrak{q}_{\xi} \rpar \exp \lpar \frac{\im}{\hbar} \mathfrak{p}_{\eta} \hat{Q} \rpar \exp \lpar -
\frac{\im}{\hbar} \mathfrak{q}_{\xi} \hat{P} \rpar ,
\ed
where the limits $R_{\mathfrak{p}} = \ell D_{\mathfrak{p}}$ and $R_{\mathfrak{q}} = \ell D_{\mathfrak{q}}$ are related to the
maximum range of each discrete spectrum $\{ \mathfrak{p}_{\eta} \}$ and $\{ \mathfrak{q}_{\xi} \}$.

\item The continuum limit $N \rightarrow \infty$ of $\hat{G}(0,0)$ reobtains the Weyl-Wigner mapping kernel
\bd
\fl \qquad \hat{\Delta}(0,0) = \int_{\mathbb{R}^{2}} \frac{d\mathfrak{p} d\mathfrak{q}}{2 \pi \hbar} \exp \lpar - 
\frac{\im}{2 \hbar} \mathfrak{p} \mathfrak{q} \rpar \exp \lpar \frac{\im}{\hbar} \mathfrak{p} \hat{Q} \rpar \exp \lpar - 
\frac{\im}{\hbar} \mathfrak{q} \hat{P} \rpar ,
\ed
or alternatively,
\be
\lb{eqa4}
\hat{\Delta}(0,0) = 2 \int_{\mathbb{R}} d \mathtt{x} \, | - \mathtt{x} \rg \lg \mathtt{x} | = 2 \hat{\mathcal{P}} .
\ee
In other words, the link with continuous parity operator $\hat{\mathcal{P}}$ becomes evident in such a case. Summarizing, the
answer of that first initial question is \texttt{yes} and particularly given by \eref{eqa4}.
\end{itemize}

For sake of completeness, it should be stressed that similar quantum representations of finite-dimensional discrete phase spaces
can also be constructed from this context \cite{Fer2011} and worked out to describe the discrete Wigner function. Exemplifying,
let us introduce the unitary operator basis \cite{Vourdas1996,Vourdas2003}
\be
\lb{eqa5}
\hat{D}(\eta,\xi) \col \om^{- \{ 2^{-1} \eta \xi \} } \hat{U}^{\eta} \hat{V}^{-\xi} \qquad \lpar \eta,\xi \in [-\ell,\ell] \rpar
\ee
which differs from $\hat{S}_{\mathrm{S}}(\eta,\xi)$ through the presence, among other things, of a specific phase here depicted
by $\om^{- \{ 2^{-1} \eta \xi \} }$. The argument showed in this phase obeys, in particular, the following rule: $2 \{ 2^{-1}
\eta \xi \} = \eta \xi + kN (\forall k \in \mathbb{Z})$, namely, $2^{-1}$ represents the multiplicative inverse of $2$ in
$\mathbb{Z}_{N}$ for $N$ odd or prime. Besides, the property
\bd
\hat{\mathsf{P}} = \frac{1}{N} \sum_{\eta,\xi=-\ell}^{\ell} \hat{D}(\eta,\xi)
\ed
exhibits a straightforward connection with the discrete parity operator, which allows us to establish the 
$\mathit{mod(N)}$-invariant unitary operator basis $\{ \hat{\Delta}(\mu,\nu) \}_{\mu,\nu=-\ell,\ldots,\ell}$ as being an unitary
transformation on $\hat{\mathsf{P}}$, that is,
\be
\lb{eqa6}
\hat{\Delta}(\mu,\nu) \col \hat{D}(\mu,\nu) \, \hat{\mathsf{P}} \, \hat{D}^{\dagger}(\mu,\nu) = \hat{D}(\mu,\nu) \hat{\Delta}(0,0)
\hat{D}^{\dagger}(\mu,\nu) .
\ee
\Eref{eqa6} can be comprehended as a further discrete counterpart of the continuous case, such that $W(\mu,\nu) \col \Tr [
\hat{\Delta}(\mu,\nu) \ro ]$ defines its respective Wigner function \cite{MR2012,WHHH2019}.\ftn{A pertinent discussion, from the
algebraic point of view, involving the effects of $\hat{S}_{\mathrm{S}}(\eta,\xi)$ and $\hat{D}(\eta,\xi)$ on the respective 
definitions of discrete Wigner function can be found in Ref. \cite{KM2005}, where, in particular, both the discrete Wigner
functions were properly calculated for the discrete vaccum coherent state with $N=23$ and illustrated by means of tridimensional
plots.}

\section{Prolegomenon for $\mathbf{SU(3)}$}
\lb{apb}

The construction method of the $\mathrm{SU(3)}$ generators follows the mathematical prescription exposed in Section 3 for $N=3$.
Firstly, let us mention that $\mathrm{SU(3)}$ is basically constituted by the eight generators
\be
\lb{eqb1apb}
\{ \hat{g} \} = \lbr \hat{\mathscr{U}}_{0,1},\hat{\mathscr{V}}_{0,1},\hat{\mathscr{W}}_{0},\hat{\mathscr{U}}_{0,2},
\hat{\mathscr{V}}_{0,2},\hat{\mathscr{U}}_{1,2},\hat{\mathscr{V}}_{1,2},\hat{\mathscr{W}}_{1} \rbr
\ee
whose expressions, explicitly written as a function of the Schwinger unitary operators, are listed in \Tref{tab2}. Furthermore,
their respective matrix representations in the basis $\{ | u_{0} \rg, | u_{1} \rg, | u_{2} \rg \}$ provide exactly the well-known
Gell-Mann matrices $\lam$'s, namely,
\brr
\lb{ext-4}
\hat{\lam}_{1} \equiv \hat{\mathscr{U}}_{0,1} = \lpar \begin{array}{ccc}
0 & 1 & 0 \\
1 & 0 & 0 \\
0 & 0 & 0 \\ 
\end{array} \rpar , \qquad
\hat{\lam}_{2} \equiv \hat{\mathscr{V}}_{0,1} = \lpar \begin{array}{ccc}
0   & - \im & 0 \\
\im &     0 & 0 \\
0   &     0 & 0 \\ 
\end{array} \rpar , \nn \\
\hat{\lam}_{3} \equiv \hat{\mathscr{W}}_{0} = \lpar \begin{array}{ccc}
1 &  0 & 0 \\
0 & -1 & 0 \\
0 &  0 & 0 \\ 
\end{array} \rpar , \qquad
\hat{\lam}_{4} \equiv \hat{\mathscr{U}}_{0,2} = \lpar \begin{array}{ccc}
0 & 0 & 1 \\
0 & 0 & 0 \\
1 & 0 & 0 \\ 
\end{array} \rpar , \nn \\
\hat{\lam}_{5} \equiv \hat{\mathscr{V}}_{0,2} = \lpar \begin{array}{ccc}
0   & 0 & - \im \\
0   & 0 & 0 \\
\im & 0 & 0 \\ 
\end{array} \rpar , \qquad
\hat{\lam}_{6} \equiv \hat{\mathscr{U}}_{1,2} = \lpar \begin{array}{ccc}
0 & 0 & 0 \\
0 & 0 & 1 \\
0 & 1 & 0 \\ 
\end{array} \rpar , \nn \\
\hat{\lam}_{7} \equiv \hat{\mathscr{V}}_{1,2} = \lpar \begin{array}{ccc}
0 & 0   & 0 \\
0 & 0   & - \im \\
0 & \im & 0 \\ 
\end{array} \rpar , \qquad
\hat{\lam}_{8} \equiv \hat{\mathscr{W}}_{1} = \lpar \begin{array}{ccc}
\frac{1}{\sqrt{3}} & 0                  & 0 \\
0                  & \frac{1}{\sqrt{3}} & 0 \\
0                  & 0                  & - \frac{2}{\sqrt{3}} \\ 
\end{array} \rpar . 
\err
%
\begin{table}[!t]
\centering
\caption{\label{tab2} Explicit constructions of the $\mathrm{SU(3)}$ generators as specific combinations of the unitary 
operators $\hat{U}$ and $\hat{V}$ -- see Equations \eref{eq14s3}-\eref{eq16s3} for $N=3$ and $\om \equiv \exp \lpar 
\frac{2 \pi \im}{3} \rpar$.}
\footnotesize
\begin{tabular}{@{}llll}
\br
{\bf Part 1.} Relations among Gell-Mann and Schwinger unitary operators \\
\mr
$\hat{\lam}_{1} = \frac{1}{3} \bigl( \hat{V} + \hat{V}^{2} + \hat{U} \hat{V} + \hat{U}^{2} \hat{V} + \om^{\ast} \hat{U} 
\hat{V}^{2} + \om \hat{U}^{2} \hat{V}^{2} \bigr)$ \\ \\
$\hat{\lam}_{2} = - \frac{\im}{3} \bigl( \hat{V} - \hat{V}^{2} + \hat{U} \hat{V} + \hat{U}^{2} \hat{V} - \om^{\ast} \hat{U} 
\hat{V}^{2} - \om \hat{U}^{2} \hat{V}^{2} \bigr)$ \\ \\
$\hat{\lam}_{3} = \frac{1}{3} \bigl[ (1-\om^{\ast}) \hat{U} + (1-\om) \hat{U}^{2} \bigr]$ \\ \\
$\hat{\lam}_{4} = \frac{1}{3} \bigl( \hat{V} + \hat{V}^{2} + \om \hat{U} \hat{V} + \om^{\ast} \hat{U}^{2} \hat{V} + \hat{U}
\hat{V}^{2} + \hat{U}^{2} \hat{V}^{2} \bigr)$ \\ \\
$\hat{\lam}_{5} = \frac{\im}{3} \bigl( \hat{V} - \hat{V}^{2} + \om \hat{U} \hat{V} + \om^{\ast} \hat{U}^{2} \hat{V} - \hat{U} 
\hat{V}^{2} - \hat{U}^{2} \hat{V}^{2} \bigr)$ \\ \\
$\hat{\lam}_{6} = \frac{1}{3} \bigl( \hat{V} + \hat{V}^{2} + \om^{\ast} \hat{U} \hat{V} + \om \hat{U}^{2} \hat{V} + \om \hat{U} 
\hat{V}^{2} + \om^{\ast} \hat{U}^{2} \hat{V}^{2} \bigr)$ \\ \\
$\hat{\lam}_{7} = - \frac{\im}{3} \bigl( \hat{V} - \hat{V}^{2} + \om^{\ast} \hat{U} \hat{V} + \om \hat{U}^{2} \hat{V} - \om
\hat{U} \hat{V}^{2} - \om^{\ast} \hat{U}^{2} \hat{V}^{2} \bigr)$ \\ \\
$\hat{\lam}_{8} = - \frac{\sqrt{3}}{3} ( \om \hat{U} + \om^{\ast} \hat{U}^{2} )$ \\
\br
{\bf Part 2.} Inverse relations for some combinations of unitary operators \\
\mr
$\hat{U} = \half \bigl[ (1-\om) \hat{\lam}_{3} - \sqrt{3} \, \om^{\ast} \hat{\lam}_{8} \bigr]$ \\ \\
$\hat{V} = \half \bigl[ \hat{\lam}_{1} + \hat{\lam}_{4} + \hat{\lam}_{6} + \im ( \hat{\lam}_{2} - \hat{\lam}_{5} + 
\hat{\lam}_{7} ) \bigr]$ \\ \\
$\hat{U}^{2} = \half \bigl[ (1-\om^{\ast}) \hat{\lam}_{3} - \sqrt{3} \, \om \hat{\lam}_{8} \bigr]$ \\ \\
$\hat{V}^{2} = \half \bigl[ \hat{\lam}_{1} + \hat{\lam}_{4} + \hat{\lam}_{6} - \im ( \hat{\lam}_{2} - \hat{\lam}_{5} + 
\hat{\lam}_{7} ) \bigr]$ \\ \\
$\hat{U} \hat{V} = \half \bigl[ \hat{\lam}_{1} + \om^{\ast} \hat{\lam}_{4} + \om \hat{\lam}_{6} + \im ( \hat{\lam}_{2} -
\om^{\ast} \hat{\lam}_{5} + \om \hat{\lam}_{7} ) \bigr]$ \\ \\
$\hat{U}^{2} \hat{V} = \half \bigl[ \hat{\lam}_{1} + \om \hat{\lam}_{4} + \om^{\ast} \hat{\lam}_{6} + \im ( \hat{\lam}_{2} - \om
\hat{\lam}_{5} + \om^{\ast} \hat{\lam}_{7} ) \bigr]$ \\ \\
$\hat{U} \hat{V}^{2} = \half \bigl[ \om \hat{\lam}_{1} + \hat{\lam}_{4} + \om^{\ast} \hat{\lam}_{6} - \im ( \om \hat{\lam}_{2} -
\hat{\lam}_{5} + \om^{\ast} \hat{\lam}_{7} ) \bigr]$ \\ \\
$\hat{U}^{2} \hat{V}^{2} = \half \bigl[ \om^{\ast} \hat{\lam}_{1} + \hat{\lam}_{4} + \om \hat{\lam}_{6} - \im ( \om^{\ast}
\hat{\lam}_{2} - \hat{\lam}_{5} + \om \hat{\lam}_{7} ) \bigr]$ \\ 
\br
\end{tabular}
\end{table}
\normalsize
As a further result, note that the product of two Gell-Mann matrices also satisfies the relation
\be
\lb{eqb2apb}
\hat{\lam}_{i} \hat{\lam}_{j} = \frac{2}{3} \delta_{ij} \hat{\mathds{I}}_{3} + \sum_{k=1}^{8} \lpar \mathscr{D}_{ijk} + \im
\mathscr{F}_{ijk} \rpar \hat{\lam}_{k} \qquad (i,j=1,\ldots,8)
\ee
with $\Tr [ \hat{\lam}_{i} \hat{\lam}_{j} ] = 2 \delta_{ij}$, for which the non-null structure constants are given by 
\cite{Ki2003}
\brr
\mathscr{F}_{123} = 1 , \nn \\ 
\mathscr{F}_{458} = \mathscr{F}_{678} = \frac{\sqrt{3}}{2} , \nn \\
\mathscr{F}_{147} = \mathscr{F}_{246} = \mathscr{F}_{257} = \mathscr{F}_{345} = - \mathscr{F}_{156} = - \mathscr{F}_{367} =
\half , \nn \\
\mathscr{D}_{118} = \mathscr{D}_{228} = \mathscr{D}_{338} = - \mathscr{D}_{888} = \frac{\sqrt{3}}{3} , \nn \\
\mathscr{D}_{448} = \mathscr{D}_{558} = \mathscr{D}_{668} = \mathscr{D}_{778} = - \frac{\sqrt{3}}{6} , \nn \\
\mathscr{D}_{146} = \mathscr{D}_{157} = \mathscr{D}_{256} = \mathscr{D}_{344} = \mathscr{D}_{355} = - \mathscr{D}_{247} = -
\mathscr{D}_{366} = - \mathscr{D}_{377} = \half . \nn
\err
An interesting discussion about Lie algebra $su(3)$ and possible applications in particle physics, emphasizing strictly its
algebraic structure, can be found in Refs. \cite{Pfeifer,Ramond,Zee,Georgi}.

According to the previous results established in Section 3 for the generators of $\mathrm{SU(N)}$, the mappings of 
$\{ \hat{g}_{i} \}_{i=1,\ldots,N^{2}-1}$ in the $N^{2}$-dimensional dual discrete phase space -- here labelled by the pair 
$(\eta,\xi)$ -- can be properly written as
\brr
\fl \qquad \bigl( \hat{\mathscr{U}}_{\alf,\bet} \bigr) (\eta,\xi) = \frac{1}{\sqrt{N}} \lbk \om^{- \eta \alf} 
\delta_{\xi, \bet-\alf}^{[N]} + \om^{- \eta \bet} \delta_{\xi, N-(\bet-\alf)}^{[N]} \rbk \om^{- \half \eta \xi} , \nn \\
\fl \qquad \bigl( \hat{\mathscr{V}}_{\alf,\bet} \bigr) (\eta,\xi) = - \frac{\im}{\sqrt{N}} \lbk \om^{- \eta \alf} 
\delta_{\xi, \bet-\alf}^{[N]} - \om^{- \eta \bet} \delta_{\xi, N-(\bet-\alf)}^{[N]} \rbk \om^{- \half \eta \xi} , \nn \\
\fl \qquad \bigl( \hat{\mathscr{W}}_{\gam} \bigr)(\eta,\xi) = \sqrt{\frac{2}{(\gam+1)(\gam+2)}} \frac{1}{\sqrt{N}} \lbk 
\sum_{\sigma = 0}^{\gam} \om^{- \eta \sigma} - (\gam + 1) \om^{-\eta (\gam + 1)} \rbk \om^{- \half \eta \xi} \, 
\delta_{\xi,0}^{[N]} . \nn
\err
In particular, these expressions represent the respective dual counterparts of Equations \eref{eq18s3}-\eref{eq20s3}, since both
the results are connected through the discrete Fourier transform
\be
\lb{eqb3apb}
\bigl( \hat{g}_{i} \bigr) (\mu,\nu) = \frac{1}{\sqrt{N}} \sum_{\eta,\xi=0}^{N-1} \om^{\mu \eta + \nu \xi} 
\om^{- \half N \Phi(\eta,\xi;N)} \bigl( \hat{g}_{i} \bigr) (\eta,\xi) .
\ee
For instance, if one considers the group $\mathrm{SU(3)}$, the dual representatives $\bigl( \hat{\lam}_{i} \bigr) (\eta,\xi)$
assume the following explicit forms:
\brr
\lb{eqb4apb}
\eqalign{
\bigl( \hat{\lam}_{1} \bigr) (\eta,\xi) = \frac{\sqrt{3}}{3} \lpar \delta_{\xi,1}^{[3]} + \om^{-\eta} \delta_{\xi,2}^{[3]} \rpar
\om^{-\half \eta \xi} , \\ 
\bigl( \hat{\lam}_{2} \bigr) (\eta,\xi) = - \im \frac{\sqrt{3}}{3} \lpar \delta_{\xi,1}^{[3]} - \om^{-\eta} \delta_{\xi,2}^{[3]}
\rpar \om^{-\half \eta \xi} , \\
\bigl( \hat{\lam}_{3} \bigr) (\eta,\xi) = \frac{\sqrt{3}}{3} \lpar 1 - \om^{-\eta} \rpar \om^{-\half \eta \xi} \, 
\delta_{\xi,0}^{[3]} \, , \\ 
\bigl( \hat{\lam}_{4} \bigr) (\eta,\xi) = \frac{\sqrt{3}}{3} \lpar \delta_{\xi,2}^{[3]} + \om^{-2\eta} \delta_{\xi,1}^{[3]} \rpar
\om^{-\half \eta \xi} , \\
\bigl( \hat{\lam}_{5} \bigr) (\eta,\xi) = - \im \frac{\sqrt{3}}{3} \lpar \delta_{\xi,2}^{[3]} - \om^{-2\eta} \delta_{\xi,1}^{[3]}
\rpar \om^{-\half \eta \xi} , \\ 
\bigl( \hat{\lam}_{6} \bigr) (\eta,\xi) = \frac{\sqrt{3}}{3} \lpar \om^{-\eta} \delta_{\xi,1}^{[3]} + \om^{-2\eta} 
\delta_{\xi,2}^{[3]} \rpar \om^{-\half \eta \xi} , \\
\bigl( \hat{\lam}_{7} \bigr) (\eta,\xi) = - \im \frac{\sqrt{3}}{3} \lpar \om^{-\eta} \delta_{\xi,1}^{[3]} - \om^{-2\eta}
\delta_{\xi,2}^{[3]} \rpar \om^{-\half \eta \xi} , \\ 
\bigl( \hat{\lam}_{8} \bigr) (\eta,\xi) = \frac{1}{3} \lpar 1 + \om^{-\eta} - 2 \om^{-2\eta} \rpar \om^{-\half \eta \xi} \,
\delta_{\xi,0}^{[3]} \, .}
\err
The desired results are a consequence of taking the discrete Fourier transform for each expression separately, that is,
\brr
\lb{eqb5apb}
\eqalign{
\bigl( \hat{\lam}_{1} \bigr) (\mu,\nu) = \frac{2}{3} \frac{\sin \lbk \lpar \mu - \half \rpar \pi \rbk}{\sin \lbk \lpar 
\mu - \half \rpar \frac{\pi}{3} \rbk} \cos \lpar \frac{2 \pi \nu}{3} \rpar , \\
\bigl( \hat{\lam}_{2} \bigr) (\mu,\nu) = \frac{2}{3} \frac{\sin \lbk \lpar \mu - \half \rpar \pi \rbk}{\sin \lbk \lpar \mu - \half
\rpar \frac{\pi}{3} \rbk} \sin \lpar \frac{2 \pi \nu}{3} \rpar , \\
\bigl( \hat{\lam}_{3} \bigr) (\mu,\nu) = \delta_{\mu,0}^{[3]} - \delta_{\mu,1}^{[3]} , \\
\bigl( \hat{\lam}_{4} \bigr) (\mu,\nu) = 2 \delta_{\mu,1}^{[3]} \cos \lpar \frac{4 \pi \nu}{3} \rpar , \\
\bigl( \hat{\lam}_{5} \bigr) (\mu,\nu) = 2 \delta_{\mu,1}^{[3]} \sin \lpar \frac{4 \pi \nu}{3} \rpar , \\
\bigl( \hat{\lam}_{6} \bigr) (\mu,\nu) = \frac{2}{3} \frac{\sin \lbk \lpar \mu - \frac{3}{2} \rpar \pi \rbk}{\sin \lbk 
\lpar \mu - \frac{3}{2} \rpar \frac{\pi}{3} \rbk} \cos \lpar \frac{2 \pi \nu}{3} \rpar , \\
\bigl( \hat{\lam}_{7} \bigr) (\mu,\nu) = \frac{2}{3} \frac{\sin \lbk \lpar \mu - \frac{3}{2} \rpar \pi \rbk}{\sin \lbk \lpar \mu -
\frac{3}{2} \rpar \frac{\pi}{3} \rbk} \sin \lpar \frac{2 \pi \nu}{3} \rpar , \\
\bigl( \hat{\lam}_{8} \bigr) (\mu,\nu) = \frac{\sqrt{3}}{3} \lpar \delta_{\mu,0}^{[3]} + \delta_{\mu,1}^{[3]} - 2 
\delta_{\mu,2}^{[3]} \rpar .} 
\err
These mapped expressions represent important parts for the evaluation of discrete $\mathrm{SU(3)}$ Wigner function.


\end{document}